\documentclass{article} 

\usepackage{graphicx} 
\usepackage[utf8]{inputenc}
\usepackage[]{amsthm} 
\usepackage{caption}
\usepackage[]{amssymb} 
\usepackage[]{amsmath,txfonts}
\usepackage{float}
\usepackage{longtable}
\usepackage{bm}
\usepackage{multirow}
\usepackage{subcaption}
\usepackage{graphicx}
\graphicspath{ {./images/} }
\usepackage[toc,page]{appendix}
\usepackage{lscape}
\usepackage{booktabs}
\usepackage{array}    
\usepackage{hyperref}
\hypersetup{
    colorlinks=true, allcolors = blue
    }

\usepackage[style=apa]{biblatex}
\usepackage{url}
\usepackage{times}
\usepackage{authblk}
\usepackage{stackengine}
\usepackage{pdflscape}

\title{Ablation Studies for Novel Treatment Effect Estimation Models}
\author[1]{Hugo Gobato Souto}
\affil[1]{\stackunder{{\stackunder{Institute of Mathematics and Computer Sciences at University of São Paulo, Brazil}{Av. Trab. São Carlense 400, 13566-590 São Carlos (SP), Brazil}}}{\stackunder{{hgsouto@usp.br}. {https://orcid.org/0000-0002-7039-0572}}}}
\author[2]{Francisco Louzada}
\affil[2]{\stackunder{{\stackunder{Institute of Mathematics and Computer Sciences at University of São Paulo, Brazil}{Av. Trab. São Carlense, 400, São Carlos, 13566-590, Brazil}}}{\stackunder{{louzada@icmc.usp.br}. {https://orcid.org/0000-0001-7815-9554}}}}
\date{}

\begin{document}

\maketitle

\begin{abstract}%
  Ablation studies are essential for understanding the contribution of individual components within complex models, yet their application in nonparametric treatment effect estimation remains limited. This paper emphasizes the importance of ablation studies by examining the Bayesian Causal Forest (BCF) model, particularly the inclusion of the estimated propensity score \(\hat{\pi}(x_i)\) intended to mitigate regularization-induced confounding (RIC). Through a partial ablation study utilizing a total of nine synthetic, we demonstrate that excluding \(\hat{\pi}(x_i)\) does not diminish the model's performance in estimating average and conditional average treatment effects or in uncertainty quantification. Moreover, omitting \(\hat{\pi}(x_i)\) reduces computational time by approximately 21\%. These findings could suggest that the BCF model's inherent flexibility suffices in adjusting for confounding without explicitly incorporating the propensity score. The study advocates for the routine use of ablation studies in treatment effect estimation to ensure model components are essential and to prevent unnecessary complexity.
\end{abstract}


\section{Introduction}

In recent years, the field of machine learning has witnessed a substantial proliferation of complex models, particularly within the realm of neural networks \parencite{Souto2024,Olivares2023,6697897,adhikari-etal-2019-rethinking,https://doi.org/10.48550/arxiv.1608.06037}. A critical aspect of model development and validation that has gained prominence is the use of \textit{ablation studies}. Ablation studies involve systematically removing or altering components of a model to assess their individual contributions to overall performance \parencite{NEURIPS2020_cdf6581c,Zeiler2014, NIPS2017_f9be311e,Wu2022}. This methodology enables researchers to dissect models, understand the significance of each component, and optimize architectures for enhanced efficiency and effectiveness.

The importance of ablation studies is well-recognized in the machine learning literature \parencite{Sheikholeslami2021,Du_2020_WACV, https://doi.org/10.48550/arxiv.1901.08644}, especially in the development of neural networks for tasks such as image recognition and natural language processing. For example, \textcite{He_2016_CVPR} utilized ablation studies to demonstrate the impact of residual connections on deep convolutional neural network performance, leading to the development of the ResNet architecture. Similarly, \textcite{Szegedy_2015_CVPR} employed ablation experiments to refine the Inception network by analyzing the contributions of various convolutional filter sizes to the network's ability to capture different levels of image features.

Despite their widespread application in general machine learning research, the practice of conducting ablation studies is noticeably lacking in the literature on treatment effect estimation, particularly with complex nonparametric models used for estimating the Average Treatment Effect (ATE) and the Conditional Average Treatment Effect (CATE). Many studies in this domain introduce sophisticated models without thoroughly investigating the necessity and influence of their individual components \parencite{Hassanpour2020Learning,pmlr-v70-shalit17a,Wager2018,https://doi.org/10.48550/arxiv.2407.07067,Hill2011, Hahn2020,Wu2022a}. This omission can be problematic, as it may lead to the adoption of unnecessarily complex models, potential overfitting, and inefficiencies that could be mitigated through careful analysis.

In this paper, we highlight the importance of ablation studies in the context of treatment effect estimation by examining the Bayesian Causal Forest (BCF) model proposed by \textcite{Hahn2020}. The BCF model is a prominent example, already used in various applied studies \parencite{Yeager2019,Yeager2022,Bail2019}, that incorporates the estimated propensity score into the baseline function to mitigate Regularization-induced Confounding (RIC). The inclusion of the propensity score is presented as a critical component for accurately estimating treatment effects in observational studies \parencite{Hahn2020}.

We challenge this assertion by conducting an ablation study focused on the use of the estimated propensity score within the BCF model. Utilizing synthetic data generated based on the designs proposed by \textcite{https://doi.org/10.48550/arxiv.2409.04874}, we assess the necessity of the propensity score in varying data-generating processes (DGPs) and sample sizes. Our findings indicate that the inclusion of the estimated propensity score does not enhance the performance of the BCF model in any of the tested scenarios. Furthermore, incorporating the propensity score results in a significant decrease in computational efficiency, with the model running approximately 21\% slower. These results underscore the critical role that ablation studies can play in the development and evaluation of models for treatment effect estimation. By systematically examining the contributions of individual model components, researchers can avoid unnecessary complexity, improve computational efficiency, and enhance the interpretability of their models.

The remainder of this paper is structured as follows: In Section \ref{Sec2}, we explore the role of the estimated propensity score in the BCF model and discuss the implications of the RIC problem. Section \ref{Sec3} provides a brief overview of the synthetic data used for our analysis and outlines the evaluation methods employed. In Section \ref{Sec4}, we present the results of our ablation study and offer a detailed analysis of the findings. Finally, Section \ref{Sec5} concludes the paper and highlights the importance of incorporating ablation studies in future research on treatment effect estimation.

\section{BCF Model Architecture} \label{Sec2}

The BCF model addresses a critical issue known as RIC, which arises in the context of high-dimensional covariate spaces and complex outcome models \parencite{Hahn2018}. Incidentally, RIC is accentuated when the outcome variable is largely determined by the covariates rather than the treatment \parencite{Hahn2018} and when there is considerable target selection (i.e., the propensity score is monotone in the prognostic function) \parencite{Hahn2020}. 

RIC refers to the bias that can occur when the regularization inherent in flexible modeling approaches inadvertently induces a correlation between the estimated treatment effect and the propensity score, potentially leading to incorrect causal inferences \parencite{Hahn2018}. The essence of RIC lies in the fact that, in observational studies with strong confounding and weak signal-to-noise ratios, the model's attempt to regularize towards simpler functions can cause the estimated treatment effect to absorb residual variation that is actually due to confounding, especially in the presence of considerable target selection. As a result, the posterior distribution of the treatment effect may be substantially influenced by the prior over the outcome model, rather than being driven by the data \parencite{Hahn2020}.

To mitigate the effects of RIC, the BCF model incorporates the estimated propensity score $\hat{\pi}(x_i)$ directly into the outcome model. By doing so, it adjusts for the treatment assignment mechanism in a manner that is integrated into the Bayesian framework of the model. The modified outcome model is specified as:
\begin{equation}
    \mathbb{E}[Y_i \mid X_i = x_i, Z_i = z_i] = \mu(x_i, \hat{\pi}(x_i)) + \tau(x_i) z_i,
    \label{eq:BCF_model}
\end{equation}
where $Y_i$ is the outcome variable, $X_i$ represents the covariates, $Z_i$ is the treatment indicator, $\mu(x_i, \hat{\pi}(x_i))$ is the prognostic function capturing the baseline outcome adjusted for the propensity score, and $\tau(x_i)$ is the heterogeneous treatment effect function. The inclusion of $\hat{\pi}(x_i)$, which is usually estimated with the standard BART model for binary outcomes, in the prognostic term $\mu(x_i, \hat{\pi}(x_i))$ serves to account for the relationship between the covariates and the treatment assignment, effectively controlling for confounding in the estimation of the baseline outcome. This adjustment is particularly important in scenarios where treatment assignment is related to the potential outcomes, such as in cases of \textit{targeted selection}, where individuals are assigned to treatment based on predicted outcomes under control \parencite{Hahn2020}.

For the estimation of $\mu(x_i, \hat{\pi}(x_i))$, and $\tau(x_i)$, the BCF model employs separate BART priors with different regularization parameters to reflect their distinct roles in the model. The prognostic function $\mu(x_i, \hat{\pi}(x_i))$ is modeled using a sum of regression trees with a relatively diffuse prior to capture potentially complex baseline relationships. Specifically, the prior settings for $\mu$ use 200 trees, a shrinkage parameter $\eta = 0.95$, and a tree depth parameter $\beta = 2$, following the defaults recommended by \textcite{Chipman2010}. Additionally, a half-Cauchy prior is placed on the scale parameter of the leaf nodes, with a prior median set to twice the marginal standard deviation of $Y$ \parencite{Hahn2020}. In contrast, the treatment effect function $\tau(x_i)$ is modeled with a stronger regularization to reflect the assumption that treatment effect heterogeneity may be more modest in magnitude and complexity. The prior for $\tau(x_i)$ utilizes 50 trees, a shrinkage parameter $\eta = 0.25$, and a depth parameter $\beta = 3$, imposing a stricter control over the flexibility of the function. A half-normal prior is placed on the scale parameter of the leaf nodes, anchoring the prior median to the marginal standard deviation of $Y$ \parencite{Hahn2020}.

By accounting for the treatment assignment mechanism within the outcome model, BCF aims to produce more reliable estimates of the ATE and CATE, particularly in observational studies where confounding is a significant concern. Nonetheless, despite these theoretical advantages, the necessity of including $\hat{\pi}(x_i)$ in the BCF model has not been thoroughly examined through ablation studies. Such studies are essential to ascertain whether the inclusion of the propensity score genuinely contributes to improved treatment effect estimation or if it introduces unnecessary complexity and computational burden. In our investigation, we aim to assess the impact of excluding $\hat{\pi}(x_i)$ from the BCF model by conducting a partial ablation study, challenging the assumption that the inclusion of the estimated propensity score is crucial for superb performance of the BCF model, even in cases of target selection and the outcome variable being largely determined by the covariates.

By systematically analyzing the model's performance, we provide insights into the practical significance of incorporating $\hat{\pi}(x_i)$ and contribute to the discourse on model simplification and efficiency in treatment effect estimation, making the improved BCF model 21\% faster than the proposed model by \textcite{Hahn2020}.

\section{Methodology} \label{Sec3}

\subsection{Considered Synthetic Data}
To evaluate the impact of including the estimated propensity score $\hat{\pi}(x_i)$ in the BCF model, we conducted a partial ablation study using synthetic data, which includes three models, namely the original BCF model (BCF ($\hat{\pi}(\mathbf{X})$)), the BCF model with the true propensity score BCF ($\pi(\mathbf{X})$), and the BCF model without the use of the estimated propensity score (BCF (no $\hat{\pi}(\mathbf{X})$)) (i.e., using $\hat{\pi}(\mathbf{X})=0.5 \ \forall \ X \in \mathbb{X}$). The data were generated inspired in the designs proposed by \textcite{https://doi.org/10.48550/arxiv.2409.04874}, yet modified to vary the extent to which the outcome variable is determined by the covariates relatively to the treatment and the propensity score is a monotone function of the prognostic function as the greater both are the greater the bias of the estimated treatment effect is due to RIC.

The different data-generating processes (DGPs) are given as:
\begin{align*}
X_i &\sim \text{Uniform}(0,1)^{5}, \\
D_i \mid X_i &\sim \text{Bernoulli}(\pi(X_i)), \\
\varepsilon_i &\sim \mathcal{N}(0,1), \\
Y_i &= b(X_i) + (D_i - 0.5) \cdot \left( \frac{X_{i,1} + X_{i,2}}{2\alpha} \right) + \varepsilon_i,
\end{align*}
where \( X_i \) is a 5-dimensional vector of covariates for individual \( i \), \( D_i \) is the binary treatment indicator, \( Y_i \) is the observed outcome, and $\alpha$ is the hyperparameter that controls the extent to which the outcome variable is determined by the covariates relatively to the treatment. Additionally, the functions \( b(X_i) \) and \(\pi(X_i) \) define the baseline main effect (a.k.a the prognostic function) and the propensity score function, respectively. Three different values of $\alpha$ are considered in this paper, namely 1, 2, and 4, leading to the average amount of times that the baseline main effect being greater than the treatment effect being approximately 3, 7 and 13 times. Remember that the negative impact of RIC on treatment effect estimation is greater as the conditional expectation of $Y$ is largely determined by $X$ rather than $D$ \parencite{Hahn2020}.

\begin{figure}[H]
\centering
\includegraphics[scale=0.25]{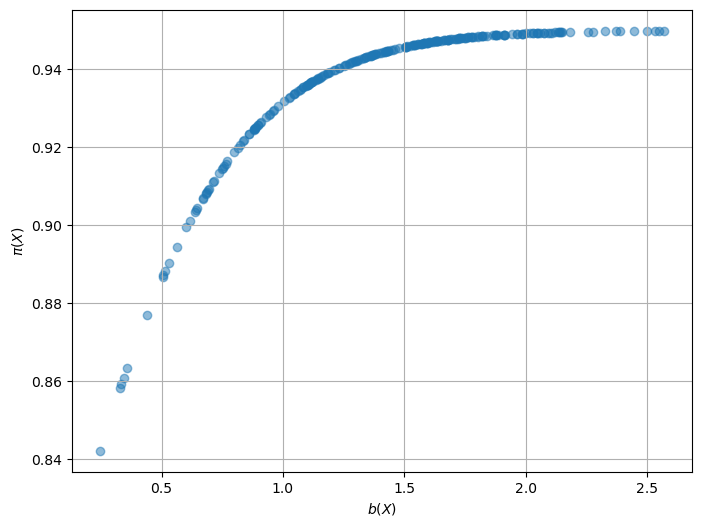}
\includegraphics[scale=0.25]{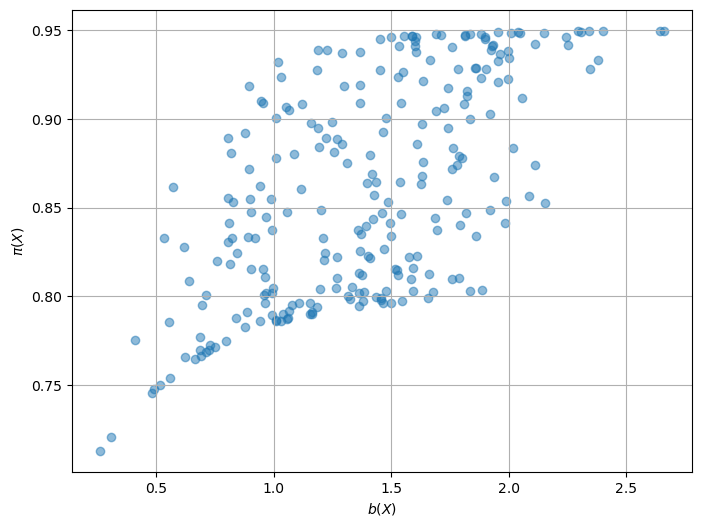}
\includegraphics[scale=0.25]{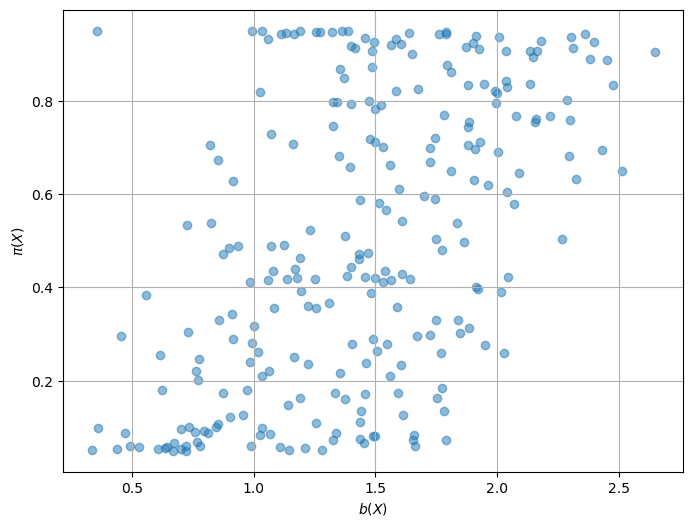}
\caption{Scatter plot of $\pi(X)$ by $b(X)$ for different configurations of $\pi(X)$}
\label{Figure1}
\centering
\end{figure}

The considered baseline and propensity scores are:
\begin{align*}
  b(X_i) &= \sin(\pi X_{i,1} X_{i,2}) + 2 (X_{i,3} - 0.5)^2 + X_{i,4} + 0.5 X_{i,5}, \\
  \pi_{\text{extreme}}(X_i) &= 0.05 + 0.9 \cdot \text{BetaCDF}_{2,4} ( \text{sigmoid}(b(X_i)) )\\
  \pi_{\text{moderate}}(X_i) &= 0.05 + 0.75 \cdot \text{BetaCDF}_{2,4}(\text{sigmoid}(b(X_i))) + 0.15 \cdot \text{BetaCDF}_{2,4}(\min(X[:, :2])), \\
  \pi_{\text{slight}}(X_i) &= 0.05 + 0.9 \cdot \text{BetaCDF}_{2,4} ( \min(X_{i,1}, X_{i,2}) ).
\end{align*}
and the considered configurations of the different DGPs are: DGP1 ($b(X); \pi_{\text{extreme}}(X)$) having an extreme target selection, DGP2 ($b(X); \pi_{\text{moderate}}(X)$) having a moderate target selection, and DGP3 ($b(X); \pi_{\text{slight}}(X)$) having a slight target selection. For each DGP, we considered the sample size of \( n = 250 \) as this was the sample size used in \textcite{Hahn2020} when presenting an example of how the use of BCF ($\hat{\pi}(\mathbf{X})$) instead of BCF (no $\hat{\pi}(\mathbf{X})$) mitigates RIC. Anew, the negative impact of RIC on treatment effect estimation is greater as the extent to which the propensity score is a monotone function of the prognostic function increases (i.e., the target selection is more pronounced) \parencite{Hahn2020}. A visual illustration of the extent to which the different $\pi(X)$ configurations are a monotone function of the prognostic function can be found in Figure \ref{Figure1}.

For each DGP, we considered the sample size of \( n = 250 \) as this was the sample size used in \textcite{Hahn2020} when presenting an example of how the use of BCF ($\hat{\pi}(\mathbf{X})$) instead of BCF (no $\hat{\pi}(\mathbf{X})$) mitigates RIC.

Nevertheless, to ensure the sparsity of this paper, we focus on the most challenging scenarios in the main text. Specifically, we present the results for $alpha=4$ of all DGPs. These DGPs were selected because are the contexts in which the use of BCF ($\hat{\pi}(\mathbf{X})$) instead of BCF (no $\hat{\pi}(\mathbf{X})$) would be the most beneficial according to \textcite{Hahn2020}. Yet, the results for $\alpha=1$ and $\alpha=2$ are provided in the Appendix \ref{AppendixA} and \ref{AppendixB} respectively, while Appendix \ref{AppendixC} presents some box plots to illustrate the results of the simulation study of this paper. Importantly, the findings across all $\alpha$ values lead to the same conclusions; thus, independently of the choice of $\alpha$ value for the main text, the findings of this ablation study would remain the same. Lastly, all simulations were repeated 100 times to account for variability due to random sampling and to provide robust estimates of the evaluation metrics.

\subsection{Evaluation Methods}

The performance of the BCF model, both with and without the inclusion of the estimated propensity score, was evaluated using metrics consistent with those used in \textcite{https://doi.org/10.48550/arxiv.2409.06593}, Root Mean Squared Error (RMSE), Mean Absolute Error (MAE), Mean Absolute Percentage Error (MAPE), Coverage (Cover), and Length (Len).

All these metrics were computed for both the Average Treatment Effect (ATE) and the Conditional Average Treatment Effect (CATE) while only RMSE and MAE are used for $\hat{\pi}(\mathbf{X})$ to show that the used propensity scores are indeed different for each model (as it could theoretically be the case that BCF ($\hat{\pi}(\mathbf{X})$) $\approx$ BCF ($\pi(\mathbf{X})$) or BCF (no $\hat{\pi}(\mathbf{X})$) if the estimated propensity score was remarkably accurate or innacurate respectively). Besides, the evaluation metrics results of the 100 different simulations of each DGP are presented in the form of their means and standard deviations as proposed by \textcite{https://doi.org/10.48550/arxiv.2409.05161}. 

Additionally, to determine whether differences in the evaluation metrics between the models with and without the estimated propensity score were statistically significant, we employed statistical tests selected using the algorithm proposed by \textcite{https://doi.org/10.48550/arxiv.2409.05161}. This algorithm provides a systematic approach to choosing appropriate statistical tests based on the characteristics of the data and the specific evaluation metrics.

\section{Results and Analysis} \label{Sec4}
Before going to the results of the specifics DGPs, it is worth mentioning that all analyses were performed using the same computational environment to maintain consistency. The computational time for fitting each model was recorded to assess the impact of including the estimated propensity score on computational efficiency. The inclusion of \( \hat{\pi}(x_i) \) increased the computational time by approximately 21\%, demonstrating a trade-off between model complexity and computational resources. As we will see below, the additional computational costs do not lead to a better performance at CATE and ATE estimation nor uncertainty quantification.

Table \ref{Table1} presents the results of the evaluation metrics for DGP1 ($b(X); \pi_{\text{extreme}}(X)$), DGP2 ($b(X); \pi_{\text{moderate}}(X)$), and DGP3 ($b(X); \pi_{\text{slight}}(X)$). It can be seen that all models perform relatively similar considering all evaluation measures and DGP1 and DGP2, while the metrics RMSE\(_{pi}\) and MAE\(_{pi}\) demonstrate that the models are indeed different (i.e., the used propensity scores are different per model). For DGP3, the performance of BCF (no $\hat{\pi}(\mathbf{X})$) is lower than the other models for the pointwise evaluation metrics, albeit statistical tests are needed to determine whether the difference in performance between BCF ($\hat{\pi}(\mathbf{X})$) and BCF (no $\hat{\pi}(\mathbf{X})$) are statistically significantly. Besides, the results are a bit contradicting to the affirmations of \textcite{Hahn2020} as we would expect that the negative impact of RIC would be more pronunced in DGP1 and DGP2, but here it appears to be present only in DGP3.

Moving to the statistical tests, Tables \ref{Table2},\ref{Table4}, and \ref{Table6} present the p-values for the statistical tests of the evaluation measures for the comparison of BCF ($\hat{\pi}(\mathbf{X})$) and BCF (no $\hat{\pi}(\mathbf{X})$). It can be seen that the used propensity scores for BCF ($\hat{\pi}(\mathbf{X})$) and BCF (no $\hat{\pi}(\mathbf{X})$) are indeed statistically different, and even without using any estimation of the propensity score, the BCF (no $\hat{\pi}(\mathbf{X})$) model achieves a performance that is statistically significantly similar to the BCF ($\hat{\pi}(\mathbf{X})$) model for all evaluation metrics. The only exception here would be the mean value of Len\(_{ATE}\) for DG1 and DGP2, yet given the similarity in coverage for these models, such statistically significant difference is not important or indicative of a superiority of one model over the other in uncertainty quantification.

Hence, it can be concluded that the estimation of the propensity score and its use is not only not necessary for the considerable performance of the BCF model in treatment effect estimation (even in cases of target selection and the outcome variable being largely determined by the covariates), but also leads to additional computational costs, increasing the amount of time needed to fit the model by roughly 21\%. Consequently, the BCF model ought to simply not estimate the propensity score function when estimating CATE and ATE (in any case not seperately as advocated by \textcite{Hahn2020} as perhaps the model can already estimate it directly thanks to its flexibility, albeit more research would be needed to explore this hypothesis), unless the researchers and/practitioners are interested in the estimation of the propensity score function.

\begin{table}[H]
\centering
\caption{Mean and Standard Deviation for Different Metrics}\label{Table1}
\begin{tabular}{llccc}
\toprule
DGP & Variable & BCF (no $\hat{\pi}(\mathbf{X})$) & BCF ($\pi(\mathbf{X})$) & BCF ($\hat{\pi}(\mathbf{X})$) \\
\midrule
DGP1 & RMSE\(_{CATE}\) & $0.157 \pm 0.119$ & $0.155 \pm 0.112$ & $0.155 \pm 0.106$ \\
& MAE\(_{CATE}\)  & $0.146 \pm 0.121$ & $0.146 \pm 0.114$ & $0.145 \pm 0.106$ \\
& MAPE\(_{CATE}\) & $1.57 \pm 1.37$   & $1.54 \pm 1.30$   & $1.54 \pm 1.24$ \\
& Cover\(_{CATE}\)& $0.992 \pm 0.0415$ & $0.996 \pm 0.0238$ & $0.998 \pm 0.0152$ \\
& Len\(_{CATE}\)  & $1.16 \pm 0.170$  & $1.10 \pm 0.153$  & $1.19 \pm 0.176$ \\
& RMSE\(_{ATE}\)  & $0.139 \pm 0.127$ & $0.139 \pm 0.120$ & $0.138 \pm 0.112$ \\
& MAE\(_{ATE}\)   & $0.139 \pm 0.127$ & $0.139 \pm 0.120$ & $0.138 \pm 0.112$ \\
& MAPE\(_{ATE}\)  & $1.10 \pm 1.00$   & $1.11 \pm 0.950$  & $1.10 \pm 0.898$ \\
& Cover\(_{ATE}\) & $0.96 \pm 0.197$  & $0.98 \pm 0.141$  & $0.99 \pm 0.1$ \\
& Len\(_{ATE}\)   & $0.869 \pm 0.132$ & $0.863 \pm 0.130$ & $0.914 \pm 0.140$ \\
& RMSE\(_{\pi}\)  & $0.438 \pm 0.00104$ & $0 \pm 0$        & $0.0446 \pm 0.00676$ \\
& MAE\(_{\pi}\)   & $0.438 \pm 0.00110$ & $0 \pm 0$        & $0.0370 \pm 0.00581$ \\
& $\text{SE}_{\text{Cover}_{CATE}}$ & $0.00351 \pm 0.00991$ & $0.00266 \pm 0.00255$ & $0.00251 \pm 0.000627$ \\
& $\text{AE}_{\text{Cover}_{CATE}}$ & $0.0521 \pm 0.0282$ & $0.0498 \pm 0.0136$ & $0.0498 \pm 0.00588$ \\
\midrule
DGP2 & RMSE\(_{CATE}\) & $0.130 \pm 0.0827$ & $0.129 \pm 0.0811$ & $0.134 \pm 0.0897$ \\
& MAE\(_{CATE}\)  & $0.119 \pm 0.0827$ & $0.117 \pm 0.0816$ & $0.122 \pm 0.0898$ \\
& MAPE\(_{CATE}\) & $1.30 \pm 0.955$   & $1.25 \pm 0.907$   & $1.29 \pm 1.01$ \\
& Cover\(_{CATE}\)& $0.998 \pm 0.00834$ & $0.997 \pm 0.0171$ & $0.997 \pm 0.0158$ \\
& Len\(_{CATE}\)  & $0.930 \pm 0.142$  & $0.901 \pm 0.132$  & $0.959 \pm 0.143$ \\
& RMSE\(_{ATE}\)  & $0.110 \pm 0.0883$ & $0.108 \pm 0.0889$ & $0.113 \pm 0.0957$ \\
& MAE\(_{ATE}\)   & $0.110 \pm 0.0883$ & $0.108 \pm 0.0889$ & $0.113 \pm 0.0957$ \\
& MAPE\(_{ATE}\)  & $0.877 \pm 0.697$  & $0.865 \pm 0.705$  & $0.904 \pm 0.757$ \\
& Cover\(_{ATE}\) & $0.95 \pm 0.219$   & $0.97 \pm 0.171$   & $0.97 \pm 0.171$ \\
& Len\(_{ATE}\)   & $0.638 \pm 0.0775$ & $0.628 \pm 0.0767$ & $0.676 \pm 0.0882$ \\
& RMSE\(_{\pi}\)  & $0.366 \pm 0.00355$ & $0 \pm 0$         & $0.0787 \pm 0.0119$ \\
& MAE\(_{\pi}\)   & $0.362 \pm 0.00357$ & $0 \pm 0$         & $0.0616 \pm 0.00974$ \\
& $\text{SE}_{\text{Cover}_{CATE}}$ & $0.00240 \pm 0.000462$ & $0.00249 \pm 0.000832$ & $0.00243 \pm 0.000349$ \\
& $\text{AE}_{\text{Cover}_{CATE}}$ & $0.0484 \pm 0.00761$ & $0.0492 \pm 0.00848$ & $0.0491 \pm 0.00498$ \\
\midrule
DGP3 & RMSE\(_{CATE}\) & $0.147 \pm 0.0988$ & $0.131 \pm 0.0874$ & $0.129 \pm 0.0828$ \\
& MAE\(_{CATE}\)  & $0.132 \pm 0.0967$ & $0.117 \pm 0.0844$ & $0.115 \pm 0.0798$ \\
& MAPE\(_{CATE}\) & $1.65 \pm 1.33$    & $1.34 \pm 1.18$    & $1.28 \pm 1.05$ \\
& Cover\(_{CATE}\)& $0.983 \pm 0.0651$ & $0.984 \pm 0.0774$ & $0.989 \pm 0.0563$ \\
& Len\(_{CATE}\)  & $0.851 \pm 0.147$  & $0.818 \pm 0.151$  & $0.835 \pm 0.146$ \\
& RMSE\(_{ATE}\)  & $0.117 \pm 0.105$  & $0.0997 \pm 0.0936$& $0.0997 \pm 0.0874$ \\
& MAE\(_{ATE}\)   & $0.117 \pm 0.105$  & $0.0997 \pm 0.0936$& $0.0997 \pm 0.0874$ \\
& MAPE\(_{ATE}\)  & $0.940 \pm 0.852$  & $0.800 \pm 0.758$  & $0.799 \pm 0.705$ \\
& Cover\(_{ATE}\) & $0.91 \pm 0.288$   & $0.94 \pm 0.239$   & $0.97 \pm 0.171$ \\
& Len\(_{ATE}\)   & $0.555 \pm 0.0589$ & $0.542 \pm 0.0618$ & $0.564 \pm 0.0669$ \\
& RMSE\(_{\pi}\)  & $0.312 \pm 0.00659$ & $0 \pm 0$         & $0.119 \pm 0.0153$ \\
& MAE\(_{\pi}\)   & $0.280 \pm 0.00822$ & $0 \pm 0$         & $0.0928 \pm 0.0115$ \\
& $\text{SE}_{\text{Cover}_{CATE}}$ & $0.00525 \pm 0.0136$ & $0.00706 \pm 0.0376$ & $0.00467 \pm 0.0190$ \\
& $\text{AE}_{\text{Cover}_{CATE}}$ & $0.0589 \pm 0.0424$ & $0.0593 \pm 0.0598$ & $0.0552 \pm 0.0405$ \\
\end{tabular}
\end{table}

\begin{landscape}
  \begin{table}[h!]
  \centering
  \caption{Statistical Test Results: p-values for Different Metrics (DGP1)}\label{Table2}
  \begin{tabular}{lccccc}
  \hline
  \textbf{Metric} & \textbf{Fligner-Policello Test} & \textbf{Mann-Whitney U Test} & \textbf{Kruskal-Wallis H Test} & \textbf{Levene's Test} & \textbf{Brown-Forsythe Test} \\
  \hline
  RMSE\(_{CATE}\) & \textit{N/A}                  & $0.7461$     & $0.7452$     & $0.5237$      & $0.6726$   \\
  MAE\(_{CATE}\)  & \textit{N/A}                  & $0.7369$     & $0.7360$     & $0.4718$      & $0.6557$   \\
  MAPE\(_{CATE}\) & \textit{N/A}                  & $0.8364$     & $0.8355$     & $0.5210$      & $0.7418$   \\
  Cover\(_{CATE}\)& $0.5375$                       & \textit{N/A} & \textit{N/A} & $0.0196$      & $0.2302$   \\
  Len\(_{CATE}\)  & \textit{N/A}                  & $0.2279$     & $0.2274$     & $0.6157$      & $0.7291$   \\
  RMSE\(_{ATE}\)  & \textit{N/A}                  & $0.7149$     & $0.7140$     & $0.4669$      & $0.6212$   \\
  MAE\(_{ATE}\)   & \textit{N/A}                  & $0.7149$     & $0.7140$     & $0.4669$      & $0.6212$   \\
  MAPE\(_{ATE}\)  & \textit{N/A}                  & $0.7204$     & $0.7195$     & $0.4805$      & $0.6350$   \\
  Cover\(_{ATE}\) & $0.2140$                      & \textit{N/A} & \textit{N/A} & $0.0062$      & $0.1759$   \\
  Len\(_{ATE}\)   & \textit{N/A}                  & $0.0273$     & $0.0272$     & $0.4672$      & $0.5624$   \\
  RMSE\(_{pi}\)   & $0$                           & \textit{N/A} & \textit{N/A} & $1.02 \times 10^{-16}$ & $3.01 \times 10^{-16}$ \\
  MAE\(_{pi}\)    & $0$                           & \textit{N/A} & \textit{N/A} & $5.23 \times 10^{-22}$ & $4.61 \times 10^{-21}$ \\
  $\text{SE}_{\text{Cover}_{ATE}}$ & $0.2140$    & \textit{N/A} & \textit{N/A} & $0.0062$      & $0.1759$   \\
  $\text{AE}_{\text{Cover}_{ATE}}$ & $0.2140$    & \textit{N/A} & \textit{N/A} & $0.0062$      & $0.1759$   \\
  $\text{SE}_{\text{Cover}_{CATE}}$ & $0.9818$   & \textit{N/A} & \textit{N/A} & $0.0393$      & $0.2880$   \\
  $\text{AE}_{\text{Cover}_{CATE}}$ & \textit{N/A} & $0.9834$   & $0.9811$     & $0.0647$      & $0.2353$   \\
  \hline
  \end{tabular}
  \end{table}
\end{landscape}

\begin{landscape}
  \begin{table}[h!]
  \centering
  \caption{Statistical Test Results: p-values for Different Metrics (DGP2)}\label{Table3}
  \begin{tabular}{lccccc}
  \hline
  \textbf{Metric} & \textbf{Fligner-Policello Test} & \textbf{Mann-Whitney U Test} & \textbf{Kruskal-Wallis H Test} & \textbf{Levene's Test} & \textbf{Brown-Forsythe Test} \\
  \hline
  RMSE\(_{CATE}\) & \textit{N/A}                  & $0.8079$     & $0.8070$     & $0.6839$      & $0.7062$   \\
  MAE\(_{CATE}\)  & \textit{N/A}                  & $0.8174$     & $0.8164$     & $0.7134$      & $0.6998$   \\
  MAPE\(_{CATE}\) & \textit{N/A}                  & $0.7554$     & $0.7545$     & $0.9119$      & $0.8223$   \\
  Cover\(_{CATE}\)& \textit{N/A}                  & $0.9639$     & $0.9613$     & $0.0883$      & $0.4083$   \\
  Len\(_{CATE}\)  & \textit{N/A}                  & $0.1416$     & $0.1413$     & $0.6724$      & $0.7567$   \\
  RMSE\(_{ATE}\)  & \textit{N/A}                  & $0.8594$     & $0.8584$     & $0.6545$      & $0.6381$   \\
  MAE\(_{ATE}\)   & \textit{N/A}                  & $0.8594$     & $0.8584$     & $0.6545$      & $0.6381$   \\
  MAPE\(_{ATE}\)  & \textit{N/A}                  & $0.8921$     & $0.8912$     & $0.6328$      & $0.6210$   \\
  Cover\(_{ATE}\) & \textit{N/A}                  & $0.4738$     & $0.4716$     & $0.1500$      & $0.4730$   \\
  Len\(_{ATE}\)   & \textit{N/A}                  & $0.0032$     & $0.0032$     & $0.5821$      & $0.7541$   \\
  RMSE\(_{pi}\)   & $0$                           & \textit{N/A} & \textit{N/A} & $7.87 \times 10^{-13}$ & $1.64 \times 10^{-12}$ \\
  MAE\(_{pi}\)    & $0$                           & \textit{N/A} & \textit{N/A} & $1.44 \times 10^{-11}$ & $1.20 \times 10^{-10}$ \\
  $\text{SE}_{\text{Cover}_{ATE}}$ & \textit{N/A} & $0.4738$     & $0.4716$     & $0.1500$      & $0.4730$   \\
  $\text{AE}_{\text{Cover}_{ATE}}$ & \textit{N/A} & $0.4738$     & $0.4716$     & $0.1500$      & $0.4730$   \\
  $\text{SE}_{\text{Cover}_{CATE}}$ & \textit{N/A} & $0.9742$   & $0.9716$     & $0.1882$      & $0.5187$   \\
  $\text{AE}_{\text{Cover}_{CATE}}$ & \textit{N/A} & $0.9742$   & $0.9716$     & $0.1077$      & $0.4296$   \\
  \hline
  \end{tabular}
  \end{table}
\end{landscape}

\begin{landscape}
  \begin{table}[h!]
  \centering
  \caption{Statistical Test Results: p-values for Different Metrics (DGP3)}\label{Table4}
  \begin{tabular}{lccccc}
  \hline
  \textbf{Metric} & \textbf{Fligner-Policello Test} & \textbf{Mann-Whitney U Test} & \textbf{Kruskal-Wallis H Test} & \textbf{Levene's Test} & \textbf{Brown-Forsythe Test} \\
  \hline
  RMSE\(_{CATE}\) & \textit{N/A}                  & $0.2423$     & $0.2418$     & $0.1148$      & $0.1858$   \\
  MAE\(_{CATE}\)  & \textit{N/A}                  & $0.2394$     & $0.2389$     & $0.0902$      & $0.1703$   \\
  MAPE\(_{CATE}\) & \textit{N/A}                  & $0.0150$     & $0.0149$     & $0.0784$      & $0.0966$   \\
  Cover\(_{CATE}\)& \textit{N/A}                  & $0.5933$     & $0.5916$     & $0.1389$      & $0.4442$   \\
  Len\(_{CATE}\)  & \textit{N/A}                  & $0.2962$     & $0.2957$     & $0.8610$      & $0.7347$   \\
  RMSE\(_{ATE}\)  & \textit{N/A}                  & $0.3487$     & $0.3481$     & $0.0854$      & $0.1670$   \\
  MAE\(_{ATE}\)   & \textit{N/A}                  & $0.3487$     & $0.3481$     & $0.0854$      & $0.1670$   \\
  MAPE\(_{ATE}\)  & \textit{N/A}                  & $0.3388$     & $0.3382$     & $0.0890$      & $0.1722$   \\
  Cover\(_{ATE}\) & $0.0845$                      & \textit{N/A} & \textit{N/A} & $0.0003$      & $0.0747$   \\
  Len\(_{ATE}\)   & \textit{N/A}                  & $0.3891$     & $0.3884$     & $0.3974$      & $0.3728$   \\
  RMSE\(_{pi}\)   & $0$                           & \textit{N/A} & \textit{N/A} & $9.50 \times 10^{-10}$ & $4.90 \times 10^{-9}$ \\
  MAE\(_{pi}\)    & $0$                           & \textit{N/A} & \textit{N/A} & $0.0052$      & $0.0062$   \\
  $\text{SE}_{\text{Cover}_{ATE}}$ & $0.0845$    & \textit{N/A} & \textit{N/A} & $0.0003$      & $0.0747$   \\
  $\text{AE}_{\text{Cover}_{ATE}}$ & $0.0845$    & \textit{N/A} & \textit{N/A} & $0.0003$      & $0.0747$   \\
  $\text{SE}_{\text{Cover}_{CATE}}$ & \textit{N/A} & $0.3510$   & $0.3497$     & $0.6722$      & $0.8041$   \\
  $\text{AE}_{\text{Cover}_{CATE}}$ & \textit{N/A} & $0.3510$   & $0.3497$     & $0.2223$      & $0.5108$   \\
  \hline
  \end{tabular}
  \end{table}
\end{landscape}

\section{Conclusion and Recommendations}  \label{Sec5}

In this paper, we have highlighted the importance of conducting ablation studies when proposing complex nonparametric models for treatment effect estimation, such as the BCF model. Ablation studies, a staple in the machine learning literature allow for a systematic assessment of individual model components to determine their necessity and impact on performance. Despite their utility, such practices are still virtually non-existent in the treatment effect estimation literature, which can lead to the adoption of unnecessarily complex models without empirical justification. This paper sheds light to this gap in the literature and aims to motivate researchers to adopt the use of ablation studies for their future papers.

Our investigation focused on the role of the estimated propensity score \(\hat{\pi}(x_i)\) within the BCF model, which is included to mitigate RIC. We conducted a partial ablation study using synthetic data generated from three different DGPs, varying the hyperparameters $\alpha$ (responsible to the extent to which the outcome variable is determined by the covariates rather than the treatment). The results of our study indicate that incorporating the estimated propensity score into the BCF model does not lead to improved performance in estimating the ATE or CATE or uncertainty quantification.  Moreover, it is observed that including \(\hat{\pi}(x_i)\) in the BCF model resulted in a substantial increase in computational time by approximately 21\%. This additional computational cost did not yield any significant benefits in terms of estimation accuracy or uncertainty quantification. 

Our findings suggest that the BCF model's flexibility could allow it to capture the necessary relationships between covariates, treatment assignment, and outcomes without explicitly including the estimated propensity score. Alternatively, our findings could indicate that despite being argued by \textcite{Hahn2020}, RIC is not an issue for nonlinear models in most cases. In any case, our results challenge the assumption that incorporating \(\hat{\pi}(x_i)\) in the model is necessary. Nevertheless, future research should explore the underlying mechanisms that enable the BCF model to adjust for confounding without explicit incorporation of \(\hat{\pi}(x_i)\) and further test whether RIC is a problem for nonparametric models.

The implications of this study are twofold. First, it emphasizes the need for ablation studies in the development and evaluation of complex models for treatment effect estimation. By systematically assessing the contributions of individual components, researchers can avoid unnecessary complexity and improve computational efficiency without compromising performance. Second, it calls into question the necessity of including the estimated propensity score in the BCF model for the purpose of treatment effect estimation, suggesting that its inclusion should be reconsidered unless there is a specific interest in estimating the propensity score itself.

In conclusion, our study underscores the critical role of ablation studies in advancing the field of causal inference. By questioning and empirically testing the necessity of model components, researchers can develop more parsimonious, efficient, and interpretable models. We advocate for the adoption of ablation studies as a standard practice in the evaluation of new models for treatment effect estimation, ensuring that each component contributes meaningfully to the model's performance.

\section*{Supplementary Research Material and Code}
The code used in this paper can be found in the following GitHub repository: (to be shared upon publication of this paper or at resquest of the reviewers)

\section*{Acknowledgments}
The authors would like to express heartfelt gratitude to the São Paulo Research Foundation (FAPESP) for their financial support through the Master's scholarship (Processo 2024/06274-0) of the first author of this paper and the associated project (Processo Vinculado 2013/07375-0).

\printbibliography

\appendix
\section{$\alpha=1$}\label{AppendixA}

\subsection{DGP1}

\begin{table}[H]
  \centering
  \caption{Mean and Standard Deviation for Different Metrics}
  \label{Table5}
  \begin{tabular}{lccc}
  \toprule
  Variable & BCF (no $\hat{\pi}(\mathbf{X})$) & BCF ($\pi(\mathbf{X})$) & BCF $\text{Oracle}$ \\
  \midrule
  RMSE\(_{CATE}\) & $0.316 \pm 0.113$ & $0.322 \pm 0.114$ & $0.336 \pm 0.114$ \\
  MAE\(_{CATE}\)  & $0.277 \pm 0.113$ & $0.281 \pm 0.114$ & $0.297 \pm 0.115$ \\
  MAPE\(_{CATE}\) & $0.709 \pm 0.296$ & $0.716 \pm 0.294$ & $0.718 \pm 0.249$ \\
  Cover\(_{CATE}\)& $0.951 \pm 0.128$ & $0.938 \pm 0.138$ & $0.944 \pm 0.123$ \\
  Len\(_{CATE}\)  & $1.31 \pm 0.195$  & $1.26 \pm 0.185$  & $1.33 \pm 0.199$ \\
  RMSE\(_{ATE}\)  & $0.244 \pm 0.145$ & $0.248 \pm 0.148$ & $0.268 \pm 0.147$ \\
  MAE\(_{ATE}\)   & $0.244 \pm 0.145$ & $0.248 \pm 0.148$ & $0.268 \pm 0.147$ \\
  MAPE\(_{ATE}\)  & $0.488 \pm 0.289$ & $0.496 \pm 0.294$ & $0.537 \pm 0.292$ \\
  Cover\(_{ATE}\) & $0.93 \pm 0.256$  & $0.93 \pm 0.256$  & $0.93 \pm 0.256$ \\
  Len\(_{ATE}\)   & $0.963 \pm 0.147$ & $0.961 \pm 0.147$ & $1.00 \pm 0.152$ \\
  RMSE\(_{\pi}\)  & $0.438 \pm 0.00107$ & $0 \pm 0$          & $0.0463 \pm 0.00840$ \\
  MAE\(_{\pi}\)   & $0.438 \pm 0.00112$ & $0 \pm 0$          & $0.0377 \pm 0.00550$ \\
  $\text{SE}_{\text{Cover}_{CATE}}$ & $0.0162 \pm 0.0652$ & $0.0191 \pm 0.0685$ & $0.0149 \pm 0.0519$ \\
  $\text{AE}_{\text{Cover}_{CATE}}$ & $0.0710 \pm 0.106$ & $0.081 \pm 0.112$  & $0.074 \pm 0.0977$ \\
  \bottomrule
  \end{tabular}
\end{table}

\begin{landscape}
  \begin{table}[h!]
  \centering
  \caption{Statistical Test Results: p-values for Different Metrics}\label{Table6}
  \begin{tabular}{lccccc}
  \hline
  \textbf{Metric} & \textbf{Fligner-Policello Test} & \textbf{Mann-Whitney U Test} & \textbf{Kruskal-Wallis H Test} & \textbf{Levene's Test} & \textbf{Brown-Forsythe Test} \\
  \hline
  RMSE\(_{CATE}\) & \textit{N/A}                  & $0.2026$     & $0.2022$     & $0.7499$      & $0.7046$   \\
  MAE\(_{CATE}\)  & \textit{N/A}                  & $0.2214$     & $0.2209$     & $0.7043$      & $0.6648$   \\
  MAPE\(_{CATE}\) & \textit{N/A}                  & $0.3653$     & $0.3647$     & $0.2097$      & $0.2962$   \\
  Cover\(_{CATE}\)& \textit{N/A}                  & $0.5045$     & $0.5036$     & $0.6300$      & $0.6918$   \\
  Len\(_{CATE}\)  & \textit{N/A}                  & \textit{N/A} & \textit{N/A} & $0.8410$      & $0.4809$   \\
  RMSE\(_{ATE}\)  & \textit{N/A}                  & $0.2288$     & $0.2284$     & $0.9392$      & $0.9505$   \\
  MAE\(_{ATE}\)   & \textit{N/A}                  & $0.2288$     & $0.2284$     & $0.9392$      & $0.9505$   \\
  MAPE\(_{ATE}\)  & \textit{N/A}                  & $0.2346$     & $0.2341$     & $0.9488$      & $0.9566$   \\
  Cover\(_{ATE}\) & \textit{N/A}                  & $1$          & $1$          & $1$           & $1$        \\
  Len\(_{ATE}\)   & \textit{N/A}                  & \textit{N/A} & \textit{N/A} & $0.7484$      & $0.0562$   \\
  RMSE\(_{pi}\)   & $0$                           & \textit{N/A} & \textit{N/A} & $4.43 \times 10^{-14}$ & $1.23 \times 10^{-12}$ \\
  MAE\(_{pi}\)    & $0$                           & \textit{N/A} & \textit{N/A} & $4.15 \times 10^{-15}$ & $3.67 \times 10^{-15}$ \\
  $\text{SE}_{\text{Cover}_{ATE}}$ & \textit{N/A} & $1$          & $1$          & $1$           & $1$        \\
  $\text{AE}_{\text{Cover}_{ATE}}$ & \textit{N/A} & $1$          & $1$          & $1$           & $1$        \\
  $\text{SE}_{\text{Cover}_{CATE}}$ & \textit{N/A} & $0.9781$   & $0.9770$     & $0.6491$      & $0.8770$   \\
  $\text{AE}_{\text{Cover}_{CATE}}$ & \textit{N/A} & $0.9781$   & $0.9770$     & $0.7381$      & $0.8179$   \\
  \hline
  \end{tabular}
  \end{table}
\end{landscape}

\subsection{DGP2}

\begin{table}[H]
  \centering
  \caption{Mean and Standard Deviation for Different Metrics}
  \label{Table7}
  \begin{tabular}{lccc}
  \toprule
  Variable & BCF (no $\hat{\pi}(\mathbf{X})$) & BCF ($\pi(\mathbf{X})$) & BCF $\text{Oracle}$ \\
  \midrule
  RMSE\(_{CATE}\) & $0.262 \pm 0.0819$ & $0.273 \pm 0.0830$ & $0.282 \pm 0.0876$ \\
  MAE\(_{CATE}\)  & $0.222 \pm 0.0783$ & $0.230 \pm 0.0796$ & $0.241 \pm 0.0847$ \\
  MAPE\(_{CATE}\) & $0.611 \pm 0.231$  & $0.612 \pm 0.214$  & $0.611 \pm 0.185$ \\
  Cover\(_{CATE}\)& $0.963 \pm 0.0820$ & $0.946 \pm 0.112$  & $0.951 \pm 0.100$ \\
  Len\(_{CATE}\)  & $1.15 \pm 0.147$   & $1.11 \pm 0.144$   & $1.16 \pm 0.160$ \\
  RMSE\(_{ATE}\)  & $0.167 \pm 0.118$  & $0.176 \pm 0.118$  & $0.194 \pm 0.124$ \\
  MAE\(_{ATE}\)   & $0.167 \pm 0.118$  & $0.176 \pm 0.118$  & $0.194 \pm 0.124$ \\
  MAPE\(_{ATE}\)  & $0.335 \pm 0.236$  & $0.353 \pm 0.237$  & $0.390 \pm 0.248$ \\
  Cover\(_{ATE}\) & $0.96 \pm 0.197$   & $0.95 \pm 0.219$   & $0.93 \pm 0.256$ \\
  Len\(_{ATE}\)   & $0.755 \pm 0.0874$ & $0.746 \pm 0.0859$ & $0.800 \pm 0.0975$ \\
  RMSE\(_{\pi}\)  & $0.366 \pm 0.00342$& $0 \pm 0$          & $0.0809 \pm 0.0116$ \\
  MAE\(_{\pi}\)   & $0.361 \pm 0.00348$& $0 \pm 0$          & $0.0634 \pm 0.00940$ \\
  $\text{SE}_{\text{Cover}_{CATE}}$ & $0.00684 \pm 0.0302$ & $0.0125 \pm 0.0500$ & $0.00994 \pm 0.0357$ \\
  $\text{AE}_{\text{Cover}_{CATE}}$ & $0.0566 \pm 0.0606$  & $0.0684 \pm 0.0891$ & $0.0653 \pm 0.0757$ \\
  \bottomrule
  \end{tabular}
  \end{table}

\begin{landscape}
    \begin{table}[h!]
    \centering
    \caption{Statistical Test Results: p-values for Different Metrics}\label{Table8}
    \begin{tabular}{lccccc}
    \hline
    \textbf{Metric} & \textbf{Fligner-Policello Test} & \textbf{Mann-Whitney U Test} & \textbf{Kruskal-Wallis H Test} & \textbf{Levene's Test} & \textbf{Brown-Forsythe Test} \\
    \hline
    RMSE\(_{CATE}\) & \textit{N/A}                  & $0.1045$     & $0.1042$     & $0.6527$      & $0.6628$   \\
    MAE\(_{CATE}\)  & \textit{N/A}                  & $0.1055$     & $0.1052$     & $0.4684$      & $0.5239$   \\
    MAPE\(_{CATE}\) & \textit{N/A}                  & $0.4110$     & $0.4103$     & $0.1419$      & $0.3431$   \\
    Cover\(_{CATE}\)& \textit{N/A}                  & $0.2767$     & $0.2761$     & $0.1303$      & $0.3421$   \\
    Len\(_{CATE}\)  & \textit{N/A}                  & \textit{N/A} & \textit{N/A} & $0.3722$      & $0.5179$   \\
    RMSE\(_{ATE}\)  & \textit{N/A}                  & $0.1234$     & $0.1231$     & $0.7828$      & $0.7922$   \\
    MAE\(_{ATE}\)   & \textit{N/A}                  & $0.1234$     & $0.1231$     & $0.7828$      & $0.7922$   \\
    MAPE\(_{ATE}\)  & \textit{N/A}                  & $0.1246$     & $0.1243$     & $0.7674$      & $0.7752$   \\
    Cover\(_{ATE}\) & \textit{N/A}                  & $0.3549$     & $0.3533$     & $0.0628$      & $0.3546$   \\
    Len\(_{ATE}\)   & \textit{N/A}                  & \textit{N/A} & \textit{N/A} & $0.2791$      & $0.0006$   \\
    RMSE\(_{pi}\)   & $0$                           & \textit{N/A} & \textit{N/A} & $2.79 \times 10^{-19}$ & $2.93 \times 10^{-19}$ \\
    MAE\(_{pi}\)    & $0$                           & \textit{N/A} & \textit{N/A} & $2.41 \times 10^{-17}$ & $8.85 \times 10^{-17}$ \\
    $\text{SE}_{\text{Cover}_{ATE}}$ & \textit{N/A} & $0.3549$     & $0.3533$     & $0.0628$      & $0.3546$   \\
    $\text{AE}_{\text{Cover}_{ATE}}$ & \textit{N/A} & $0.3549$     & $0.3533$     & $0.0628$      & $0.3546$   \\
    $\text{SE}_{\text{Cover}_{CATE}}$ & \textit{N/A} & $0.9162$   & $0.9152$     & $0.2379$      & $0.4905$   \\
    $\text{AE}_{\text{Cover}_{CATE}}$ & \textit{N/A} & $0.9162$   & $0.9152$     & $0.0598$      & $0.2528$   \\
    \hline
    \end{tabular}
    \end{table}
\end{landscape}

\subsection{DGP3}

\begin{table}[H]
  \centering
  \caption{Mean and Standard Deviation for Different Metrics}
  \label{Table9}
  \begin{tabular}{lccc}
  \toprule
  Variable & BCF (no $\hat{\pi}(\mathbf{X})$) & BCF ($\pi(\mathbf{X})$) & BCF $\text{Oracle}$ \\
  \midrule
  RMSE\(_{CATE}\) & $0.249 \pm 0.0764$ & $0.254 \pm 0.0767$ & $0.268 \pm 0.0827$ \\
  MAE\(_{CATE}\)  & $0.208 \pm 0.0705$ & $0.212 \pm 0.0719$ & $0.225 \pm 0.0790$ \\
  MAPE\(_{CATE}\) & $0.707 \pm 0.303$  & $0.629 \pm 0.227$  & $0.628 \pm 0.200$ \\
  Cover\(_{CATE}\)& $0.949 \pm 0.0906$ & $0.940 \pm 0.122$  & $0.930 \pm 0.136$ \\
  Len\(_{CATE}\)  & $1.05 \pm 0.134$   & $1.01 \pm 0.128$   & $1.02 \pm 0.132$ \\
  RMSE\(_{ATE}\)  & $0.143 \pm 0.107$  & $0.145 \pm 0.109$  & $0.167 \pm 0.116$ \\
  MAE\(_{ATE}\)   & $0.143 \pm 0.107$  & $0.145 \pm 0.109$  & $0.167 \pm 0.116$ \\
  MAPE\(_{ATE}\)  & $0.287 \pm 0.214$  & $0.291 \pm 0.217$  & $0.333 \pm 0.231$ \\
  Cover\(_{ATE}\) & $0.88 \pm 0.327$   & $0.94 \pm 0.239$   & $0.9 \pm 0.302$ \\
  Len\(_{ATE}\)   & $0.613 \pm 0.0480$ & $0.644 \pm 0.0551$ & $0.667 \pm 0.0631$ \\
  RMSE\(_{\pi}\)  & $0.311 \pm 0.00825$& $0 \pm 0$          & $0.122 \pm 0.0152$ \\
  MAE\(_{\pi}\)   & $0.279 \pm 0.0100$ & $0 \pm 0$          & $0.0949 \pm 0.0117$ \\
  $\text{SE}_{\text{Cover}_{CATE}}$ & $0.00813 \pm 0.0279$ & $0.0147 \pm 0.0688$ & $0.0187 \pm 0.0780$ \\
  $\text{AE}_{\text{Cover}_{CATE}}$ & $0.058 \pm 0.0694$   & $0.0660 \pm 0.102$  & $0.0732 \pm 0.116$ \\
  \bottomrule
  \end{tabular}
\end{table}

\begin{landscape}
  \begin{table}[h!]
  \centering
  \caption{Statistical Test Results: p-values for Different Metrics}\label{Table10}
  \begin{tabular}{lccccc}
  \hline
  \textbf{Metric} & \textbf{Fligner-Policello Test} & \textbf{Mann-Whitney U Test} & \textbf{Kruskal-Wallis H Test} & \textbf{Levene's Test} & \textbf{Brown-Forsythe Test} \\
  \hline
  RMSE\(_{CATE}\) & \textit{N/A}                  & $0.0631$     & $0.0630$     & $0.5324$      & $0.4521$   \\
  MAE\(_{CATE}\)  & \textit{N/A}                  & $0.0693$     & $0.0691$     & $0.3840$      & $0.3522$   \\
  MAPE\(_{CATE}\) & $0.2361$                      & \textit{N/A} & \textit{N/A} & $3.82 \times 10^{-5}$ & $0.0015$   \\
  Cover\(_{CATE}\)& \textit{N/A}                  & $0.3955$     & $0.3948$     & $0.0607$      & $0.2620$   \\
  Len\(_{CATE}\)  & \textit{N/A}                  & $0.1755$     & $0.1751$     & $0.7868$      & $0.7249$   \\
  RMSE\(_{ATE}\)  & \textit{N/A}                  & $0.1463$     & $0.1460$     & $0.3762$      & $0.4473$   \\
  MAE\(_{ATE}\)   & \textit{N/A}                  & $0.1463$     & $0.1460$     & $0.3762$      & $0.4473$   \\
  MAPE\(_{ATE}\)  & \textit{N/A}                  & $0.1470$     & $0.1467$     & $0.4108$      & $0.4746$   \\
  Cover\(_{ATE}\) & \textit{N/A}                  & $0.6537$     & $0.6521$     & $0.3684$      & $0.6532$   \\
  Len\(_{ATE}\)   & $3.15 \times 10^{-13}$        & \textit{N/A} & \textit{N/A} & $0.0092$      & $0.0101$   \\
  RMSE\(_{pi}\)   & $0$                           & \textit{N/A} & \textit{N/A} & $1.01 \times 10^{-6}$ & $3.88 \times 10^{-6}$ \\
  MAE\(_{pi}\)    & \textit{N/A}                  & $2.56 \times 10^{-34}$ & $2.52 \times 10^{-34}$ & $0.1587$     & $0.1985$   \\
  $\text{SE}_{\text{Cover}_{ATE}}$ & \textit{N/A} & $0.6537$     & $0.6521$     & $0.3684$      & $0.6532$   \\
  $\text{AE}_{\text{Cover}_{ATE}}$ & \textit{N/A} & $0.6537$     & $0.6521$     & $0.3684$      & $0.6532$   \\
  $\text{SE}_{\text{Cover}_{CATE}}$ & $0.7346$  & \textit{N/A} & \textit{N/A} & $0.0188$      & $0.2043$   \\
  $\text{AE}_{\text{Cover}_{CATE}}$ & $0.7346$  & \textit{N/A} & \textit{N/A} & $0.0434$      & $0.2371$   \\
  \hline
  \end{tabular}
  \end{table}
  \end{landscape}

\section{$\alpha=2$}\label{AppendixB}

\subsection{DGP1}

\begin{table}[H]
  \centering
  \caption{Mean and Standard Deviation for Different Metrics}
  \label{Table11}
  \begin{tabular}{lccc}
  \toprule
  Variable & BCF (no $\hat{\pi}(\mathbf{X})$) & BCF ($\pi(\mathbf{X})$) & BCF $\text{Oracle}$ \\
  \midrule
  RMSE\(_{CATE}\) & $0.189 \pm 0.0795$ & $0.193 \pm 0.0831$ & $0.192 \pm 0.0930$ \\
  MAE\(_{CATE}\)  & $0.169 \pm 0.0813$ & $0.173 \pm 0.0855$ & $0.172 \pm 0.0955$ \\
  MAPE\(_{CATE}\) & $0.863 \pm 0.456$  & $0.870 \pm 0.447$  & $0.862 \pm 0.490$ \\
  Cover\(_{CATE}\)& $0.997 \pm 0.0148$ & $0.993 \pm 0.0361$ & $0.993 \pm 0.0569$ \\
  Len\(_{CATE}\)  & $1.18 \pm 0.164$   & $1.13 \pm 0.157$   & $1.22 \pm 0.164$ \\
  RMSE\(_{ATE}\)  & $0.152 \pm 0.0981$ & $0.157 \pm 0.102$  & $0.155 \pm 0.111$ \\
  MAE\(_{ATE}\)   & $0.152 \pm 0.0981$ & $0.157 \pm 0.102$  & $0.155 \pm 0.111$ \\
  MAPE\(_{ATE}\)  & $0.605 \pm 0.388$  & $0.625 \pm 0.402$  & $0.617 \pm 0.438$ \\
  Cover\(_{ATE}\) & $1.00 \pm 0.000$   & $0.98 \pm 0.141$   & $0.99 \pm 0.1$ \\
  Len\(_{ATE}\)   & $0.885 \pm 0.127$  & $0.871 \pm 0.127$  & $0.933 \pm 0.133$ \\
  RMSE\(_{\pi}\)  & $0.438 \pm 0.00108$ & $0 \pm 0$         & $0.0441 \pm 0.00648$ \\
  MAE\(_{\pi}\)   & $0.438 \pm 0.00113$ & $0 \pm 0$         & $0.0361 \pm 0.00524$ \\
  $\text{SE}_{\text{Cover}_{CATE}}$ & $0.00239 \pm 0.000436$ & $0.00313 \pm 0.00711$ & $0.00501 \pm 0.0258$ \\
  $\text{AE}_{\text{Cover}_{CATE}}$ & $0.0485 \pm 0.00627$  & $0.0507 \pm 0.0238$  & $0.0538 \pm 0.0462$ \\
  \bottomrule
  \end{tabular}
\end{table}

\begin{landscape}
  \begin{table}[h!]
  \centering
  \caption{Statistical Test Results: p-values for Different Metrics}\label{Table12}
  \begin{tabular}{lccccc}
  \hline
  \textbf{Metric} & \textbf{Fligner-Policello Test} & \textbf{Mann-Whitney U Test} & \textbf{Kruskal-Wallis H Test} & \textbf{Levene's Test} & \textbf{Brown-Forsythe Test} \\
  \hline
  RMSE\(_{CATE}\) & \textit{N/A}                  & $0.9095$     & $0.9086$     & $0.1192$      & $0.2987$   \\
  MAE\(_{CATE}\)  & \textit{N/A}                  & $0.8902$     & $0.8892$     & $0.1123$      & $0.3382$   \\
  MAPE\(_{CATE}\) & \textit{N/A}                  & $0.5650$     & $0.5642$     & $0.3988$      & $0.5722$   \\
  Cover\(_{CATE}\)& \textit{N/A}                  & $0.8533$     & $0.8514$     & $0.1769$      & $0.4842$   \\
  Len\(_{CATE}\)  & \textit{N/A}                  & $0.1505$     & $0.1501$     & $0.7967$      & $0.7875$   \\
  RMSE\(_{ATE}\)  & \textit{N/A}                  & $0.9659$     & $0.9649$     & $0.1877$      & $0.3394$   \\
  MAE\(_{ATE}\)   & \textit{N/A}                  & $0.9659$     & $0.9649$     & $0.1877$      & $0.3394$   \\
  MAPE\(_{ATE}\)  & \textit{N/A}                  & $0.9854$     & $0.9844$     & $0.1932$      & $0.3391$   \\
  Cover\(_{ATE}\) & \textit{N/A}                  & $0.5656$     & $0.5617$     & $0.2464$      & $0.5630$   \\
  Len\(_{ATE}\)   & \textit{N/A}                  & $0.0050$     & $0.0050$     & $0.7983$      & $0.7816$   \\
  RMSE\(_{pi}\)   & $0$                           & \textit{N/A} & \textit{N/A} & $8.04 \times 10^{-17}$ & $8.32 \times 10^{-17}$ \\
  MAE\(_{pi}\)    & $0$                           & \textit{N/A} & \textit{N/A} & $3.75 \times 10^{-20}$ & $1.57 \times 10^{-19}$ \\
  $\text{SE}_{\text{Cover}_{ATE}}$ & \textit{N/A} & $0.5656$     & $0.5617$     & $0.2464$      & $0.5630$   \\
  $\text{AE}_{\text{Cover}_{ATE}}$ & \textit{N/A} & $0.5656$     & $0.5617$     & $0.2464$      & $0.5630$   \\
  $\text{SE}_{\text{Cover}_{CATE}}$ & \textit{N/A} & $0.7713$   & $0.7695$     & $0.0539$      & $0.3265$   \\
  $\text{AE}_{\text{Cover}_{CATE}}$ & \textit{N/A} & $0.7713$   & $0.7695$     & $0.1676$      & $0.4002$   \\
  \hline
  \end{tabular}
  \end{table}
\end{landscape}

\subsection{DGP2}

\begin{table}[H]
  \centering
  \caption{Mean and Standard Deviation for Different Metrics}
  \label{Table13}
  \begin{tabular}{lccc}
  \toprule
  Variable & BCF (no $\hat{\pi}(\mathbf{X})$) & BCF ($\pi(\mathbf{X})$) & BCF $\text{Oracle}$ \\
  \midrule
  RMSE\(_{CATE}\) & $0.182 \pm 0.0750$ & $0.187 \pm 0.0750$ & $0.192 \pm 0.0799$ \\
  MAE\(_{CATE}\)  & $0.161 \pm 0.0747$ & $0.165 \pm 0.0755$ & $0.171 \pm 0.0799$ \\
  MAPE\(_{CATE}\) & $0.830 \pm 0.415$  & $0.829 \pm 0.395$  & $0.829 \pm 0.387$ \\
  Cover\(_{CATE}\)& $0.982 \pm 0.0772$ & $0.973 \pm 0.0890$ & $0.974 \pm 0.0702$ \\
  Len\(_{CATE}\)  & $0.984 \pm 0.159$  & $0.952 \pm 0.141$  & $0.988 \pm 0.155$ \\
  RMSE\(_{ATE}\)  & $0.142 \pm 0.0930$ & $0.146 \pm 0.0943$ & $0.154 \pm 0.0967$ \\
  MAE\(_{ATE}\)   & $0.142 \pm 0.0930$ & $0.146 \pm 0.0943$ & $0.154 \pm 0.0967$ \\
  MAPE\(_{ATE}\)  & $0.569 \pm 0.374$  & $0.587 \pm 0.379$  & $0.620 \pm 0.388$ \\
  Cover\(_{ATE}\) & $0.94 \pm 0.239$   & $0.94 \pm 0.239$   & $0.92 \pm 0.273$ \\
  Len\(_{ATE}\)   & $0.665 \pm 0.0916$ & $0.662 \pm 0.0792$ & $0.696 \pm 0.0838$ \\
  RMSE\(_{\pi}\)  & $0.366 \pm 0.00326$& $0 \pm 0$          & $0.0792 \pm 0.0106$ \\
  MAE\(_{\pi}\)   & $0.361 \pm 0.00328$& $0 \pm 0$          & $0.0624 \pm 0.00847$ \\
  $\text{SE}_{\text{Cover}_{CATE}}$ & $0.00692 \pm 0.0352$ & $0.00838 \pm 0.0370$ & $0.00545 \pm 0.0195$ \\
  $\text{AE}_{\text{Cover}_{CATE}}$ & $0.0561 \pm 0.0617$  & $0.0598 \pm 0.0697$  & $0.0558 \pm 0.0486$ \\
  \bottomrule
  \end{tabular}
  \end{table}

\begin{landscape}
  \begin{table}[h!]
    \centering
    \caption{Statistical Test Results: p-values for Different Metrics}\label{Table14}
    \begin{tabular}{lccccc}
    \hline
    \textbf{Metric} & \textbf{Fligner-Policello Test} & \textbf{Mann-Whitney U Test} & \textbf{Kruskal-Wallis H Test} & \textbf{Levene's Test} & \textbf{Brown-Forsythe Test} \\
    \hline
    RMSE\(_{CATE}\) & \textit{N/A}                  & $0.4151$     & $0.4144$     & $0.2111$      & $0.2886$   \\
    MAE\(_{CATE}\)  & \textit{N/A}                  & $0.4068$     & $0.4061$     & $0.2118$      & $0.3190$   \\
    MAPE\(_{CATE}\) & \textit{N/A}                  & $0.9134$     & $0.9124$     & $0.8095$      & $0.7188$   \\
    Cover\(_{CATE}\)& \textit{N/A}                  & $0.4644$     & $0.4635$     & $0.2499$      & $0.4329$   \\
    Len\(_{CATE}\)  & \textit{N/A}                  & $0.8921$     & $0.8912$     & $0.7035$      & $0.7675$   \\
    RMSE\(_{ATE}\)  & \textit{N/A}                  & $0.3400$     & $0.3394$     & $0.6483$      & $0.6647$   \\
    MAE\(_{ATE}\)   & \textit{N/A}                  & $0.3400$     & $0.3394$     & $0.6483$      & $0.6647$   \\
    MAPE\(_{ATE}\)  & \textit{N/A}                  & $0.3375$     & $0.3369$     & $0.6695$      & $0.6765$   \\
    Cover\(_{ATE}\) & \textit{N/A}                  & $0.5822$     & $0.5803$     & $0.2697$      & $0.5816$   \\
    Len\(_{ATE}\)   & \textit{N/A}                  & $0.0098$     & $0.0097$     & $0.4740$      & $0.5929$   \\
    RMSE\(_{pi}\)   & $0$                           & \textit{N/A} & \textit{N/A} & $2.66 \times 10^{-16}$ & $3.03 \times 10^{-16}$ \\
    MAE\(_{pi}\)    & $0$                           & \textit{N/A} & \textit{N/A} & $1.54 \times 10^{-14}$ & $2.90 \times 10^{-14}$ \\
    $\text{SE}_{\text{Cover}_{ATE}}$ & \textit{N/A} & $0.5822$     & $0.5803$     & $0.2697$      & $0.5816$   \\
    $\text{AE}_{\text{Cover}_{ATE}}$ & \textit{N/A} & $0.5822$     & $0.5803$     & $0.2697$      & $0.5816$   \\
    $\text{SE}_{\text{Cover}_{CATE}}$ & \textit{N/A} & $0.2862$   & $0.2855$     & $0.4230$      & $0.7302$   \\
    $\text{AE}_{\text{Cover}_{CATE}}$ & \textit{N/A} & $0.2862$   & $0.2855$     & $0.9172$      & $0.8185$   \\
    \hline
    \end{tabular}
  \end{table}
\end{landscape}

\subsection{DGP3}

\begin{table}[H]
  \centering
  \caption{Mean and Standard Deviation for Different Metrics}
  \label{Table15}
  \begin{tabular}{lccc}
  \toprule
  Variable & BCF (no $\hat{\pi}(\mathbf{X})$) & BCF ($\pi(\mathbf{X})$) & BCF $\text{Oracle}$ \\
  \midrule
  RMSE\(_{CATE}\) & $0.175 \pm 0.0850$ & $0.168 \pm 0.0730$ & $0.174 \pm 0.0738$ \\
  MAE\(_{CATE}\)  & $0.150 \pm 0.0832$ & $0.144 \pm 0.0716$ & $0.150 \pm 0.0734$ \\
  MAPE\(_{CATE}\) & $0.982 \pm 0.591$  & $0.830 \pm 0.466$  & $0.812 \pm 0.458$ \\
  Cover\(_{CATE}\)& $0.973 \pm 0.0993$ & $0.968 \pm 0.107$  & $0.969 \pm 0.113$ \\
  Len\(_{CATE}\)  & $0.917 \pm 0.139$  & $0.878 \pm 0.134$  & $0.883 \pm 0.133$ \\
  RMSE\(_{ATE}\)  & $0.120 \pm 0.0999$ & $0.113 \pm 0.0886$ & $0.119 \pm 0.0930$ \\
  MAE\(_{ATE}\)   & $0.120 \pm 0.0999$ & $0.113 \pm 0.0886$ & $0.119 \pm 0.0930$ \\
  MAPE\(_{ATE}\)  & $0.482 \pm 0.403$  & $0.456 \pm 0.358$  & $0.480 \pm 0.376$ \\
  Cover\(_{ATE}\) & $0.91 \pm 0.288$   & $0.93 \pm 0.256$   & $0.93 \pm 0.256$ \\
  Len\(_{ATE}\)   & $0.584 \pm 0.0467$ & $0.584 \pm 0.0542$ & $0.596 \pm 0.0539$ \\
  RMSE\(_{\pi}\)  & $0.311 \pm 0.00772$& $0 \pm 0$          & $0.121 \pm 0.0125$ \\
  MAE\(_{\pi}\)   & $0.279 \pm 0.00949$  & $0 \pm 0$          & $0.0942 \pm 0.00941$ \\
  $\text{SE}_{\text{Cover}_{CATE}}$ & $0.0103 \pm 0.0412$ & $0.0116 \pm 0.0443$ & $0.0129 \pm 0.0687$ \\
  $\text{AE}_{\text{Cover}_{CATE}}$ & $0.0652 \pm 0.0781$ & $0.0678 \pm 0.0839$ & $0.0668 \pm 0.0922$ \\
  \bottomrule
  \end{tabular}
\end{table}

\begin{landscape}
  \begin{table}[h!]
  \centering
  \caption{Statistical Test Results: p-values for Different Metrics}\label{Table16}
  \begin{tabular}{lccccc}
  \hline
  \textbf{Metric} & \textbf{Fligner-Policello Test} & \textbf{Mann-Whitney U Test} & \textbf{Kruskal-Wallis H Test} & \textbf{Levene's Test} & \textbf{Brown-Forsythe Test} \\
  \hline
  RMSE\(_{CATE}\) & \textit{N/A}                  & $0.6788$     & $0.6779$     & $0.3048$      & $0.4950$   \\
  MAE\(_{CATE}\)  & \textit{N/A}                  & $0.5850$     & $0.5842$     & $0.3460$      & $0.5840$   \\
  MAPE\(_{CATE}\) & $0.0466$                      & \textit{N/A} & \textit{N/A} & $0.0133$      & $0.0160$   \\
  Cover\(_{CATE}\)& \textit{N/A}                  & $0.7063$     & $0.7051$     & $0.6158$      & $0.7657$   \\
  Len\(_{CATE}\)  & \textit{N/A}                  & $0.0594$     & $0.0593$     & $0.6877$      & $0.6052$   \\
  RMSE\(_{ATE}\)  & \textit{N/A}                  & $0.8441$     & $0.8431$     & $0.7198$      & $0.9140$   \\
  MAE\(_{ATE}\)   & \textit{N/A}                  & $0.8441$     & $0.8431$     & $0.7198$      & $0.9140$   \\
  MAPE\(_{ATE}\)  & \textit{N/A}                  & $0.8345$     & $0.8336$     & $0.7323$      & $0.8997$   \\
  Cover\(_{ATE}\) & \textit{N/A}                  & $0.6049$     & $0.6031$     & $0.2993$      & $0.6043$   \\
  Len\(_{ATE}\)   & $0.1894$                      & \textit{N/A} & \textit{N/A} & $0.0355$      & $0.0578$   \\
  RMSE\(_{pi}\)   & $0$                           & \textit{N/A} & \textit{N/A} & $9.39 \times 10^{-7}$ & $8.86 \times 10^{-7}$ \\
  MAE\(_{pi}\)    & \textit{N/A}                  & $7.30 \times 10^{-199}$ & \textit{N/A} & $0.9324$     & \textit{N/A}   \\
  $\text{SE}_{\text{Cover}_{ATE}}$ & \textit{N/A} & $0.6049$     & $0.6031$     & $0.2993$      & $0.6043$   \\
  $\text{AE}_{\text{Cover}_{ATE}}$ & \textit{N/A} & $0.6049$     & $0.6031$     & $0.2993$      & $0.6043$   \\
  $\text{SE}_{\text{Cover}_{CATE}}$ & \textit{N/A} & $0.9701$   & $0.9688$     & $0.5377$      & $0.7445$   \\
  $\text{AE}_{\text{Cover}_{CATE}}$ & \textit{N/A} & $0.9701$   & $0.9688$     & $0.7760$      & $0.8697$   \\
  \hline
  \end{tabular}
  \end{table}
\end{landscape}

\begin{landscape}
  \section{Boxplots of Performance Measures}\label{AppendixC}
  \subsection{DGP1}
  \begin{figure}[H]
  \centering
  \includegraphics[scale=0.5]{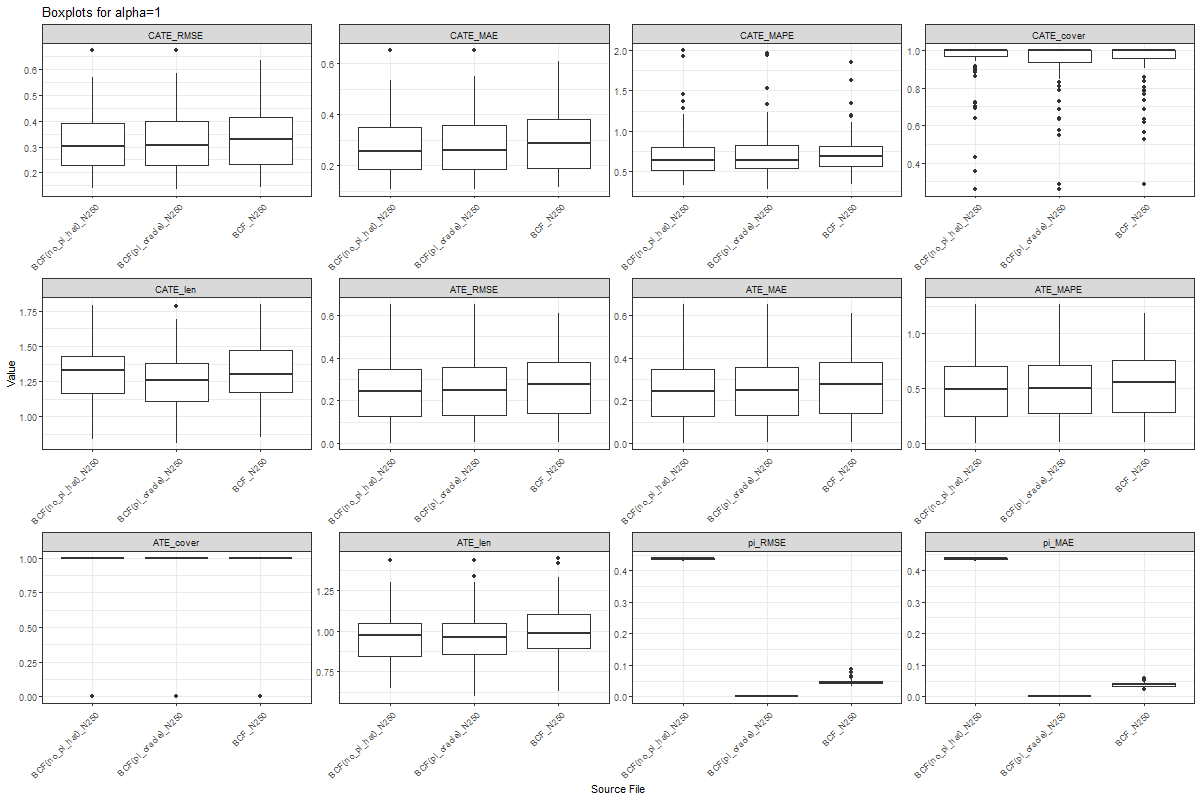}
  \caption{Box Plots for $\alpha=1$}
  \label{Figure2}
  \centering
  \end{figure}
  
  \begin{figure}[H]
  \centering
  \includegraphics[scale=0.5]{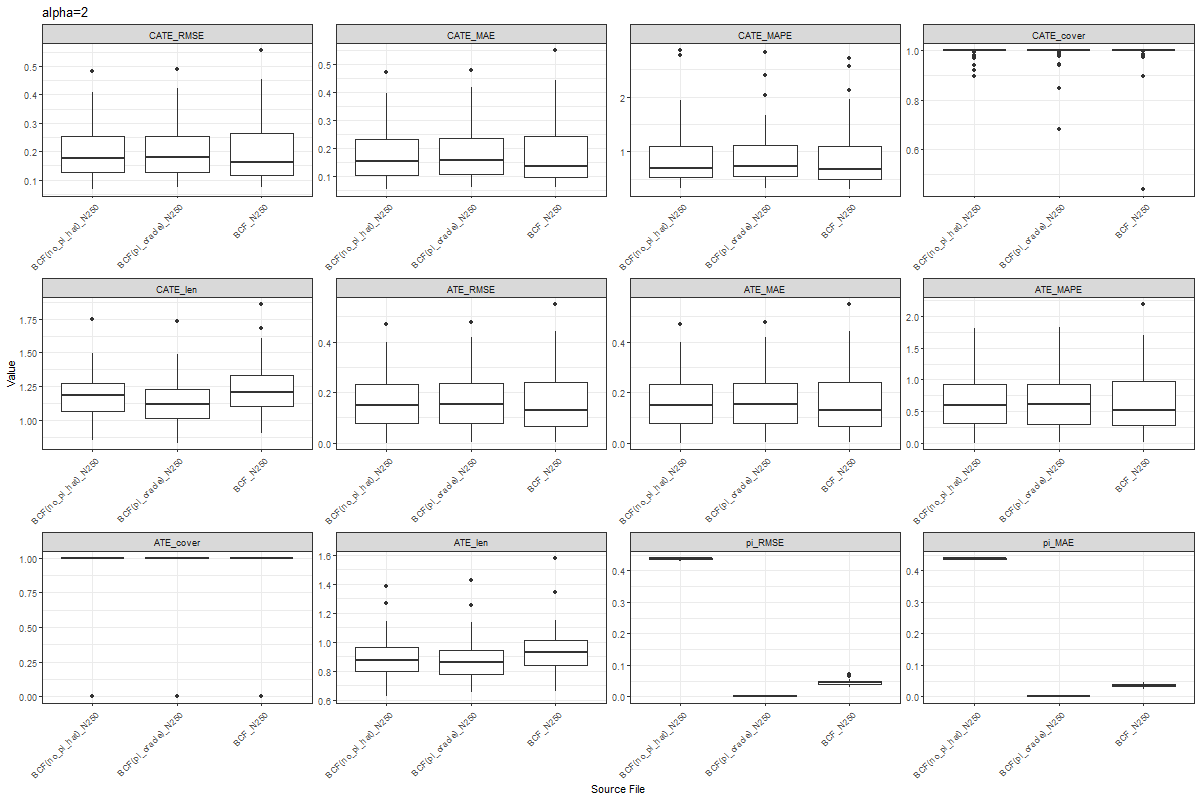}
  \caption{Box Plots for $\alpha=2$}
  \label{Figure3}
  \centering
  \end{figure}
  
  \begin{figure}[H]
  \centering
  \includegraphics[scale=0.5]{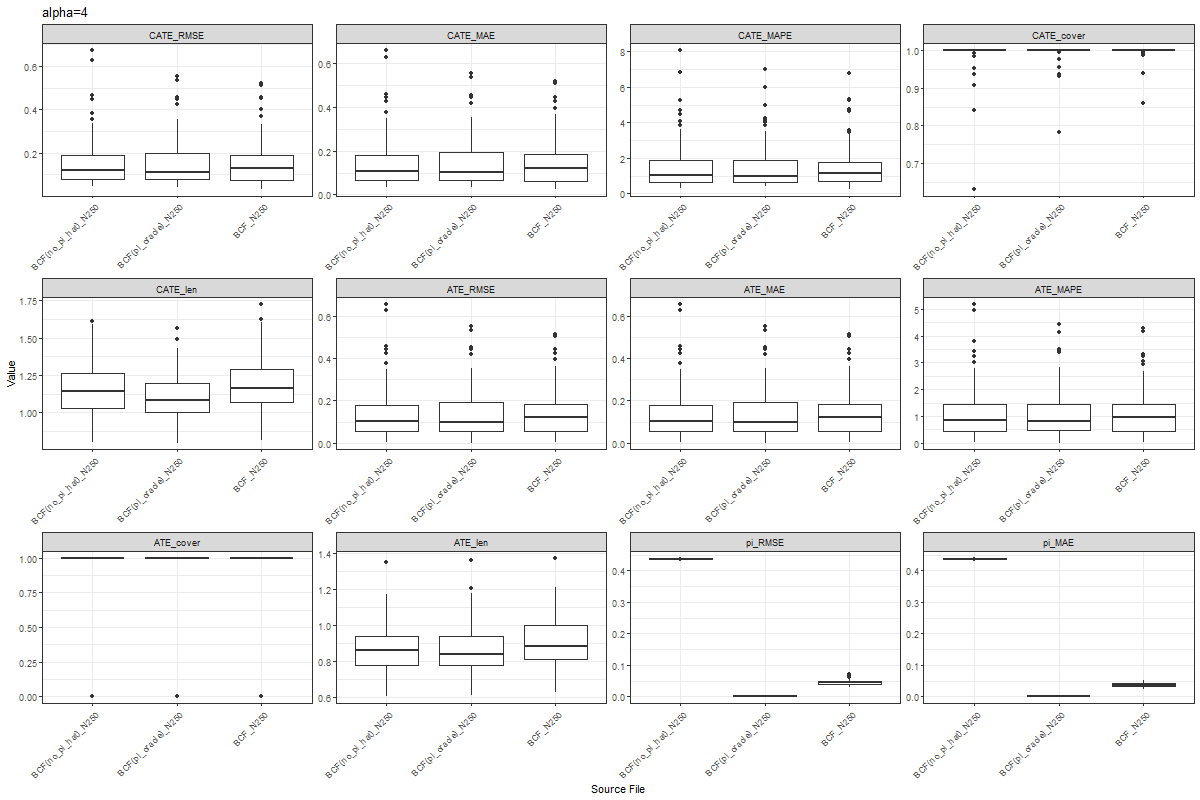}
  \caption{Box Plots for $\alpha=4$}
  \label{Figure4}
  \centering
  \end{figure}
  
  \subsection{DGP2}
  \begin{figure}[H]
  \centering
  \includegraphics[scale=0.5]{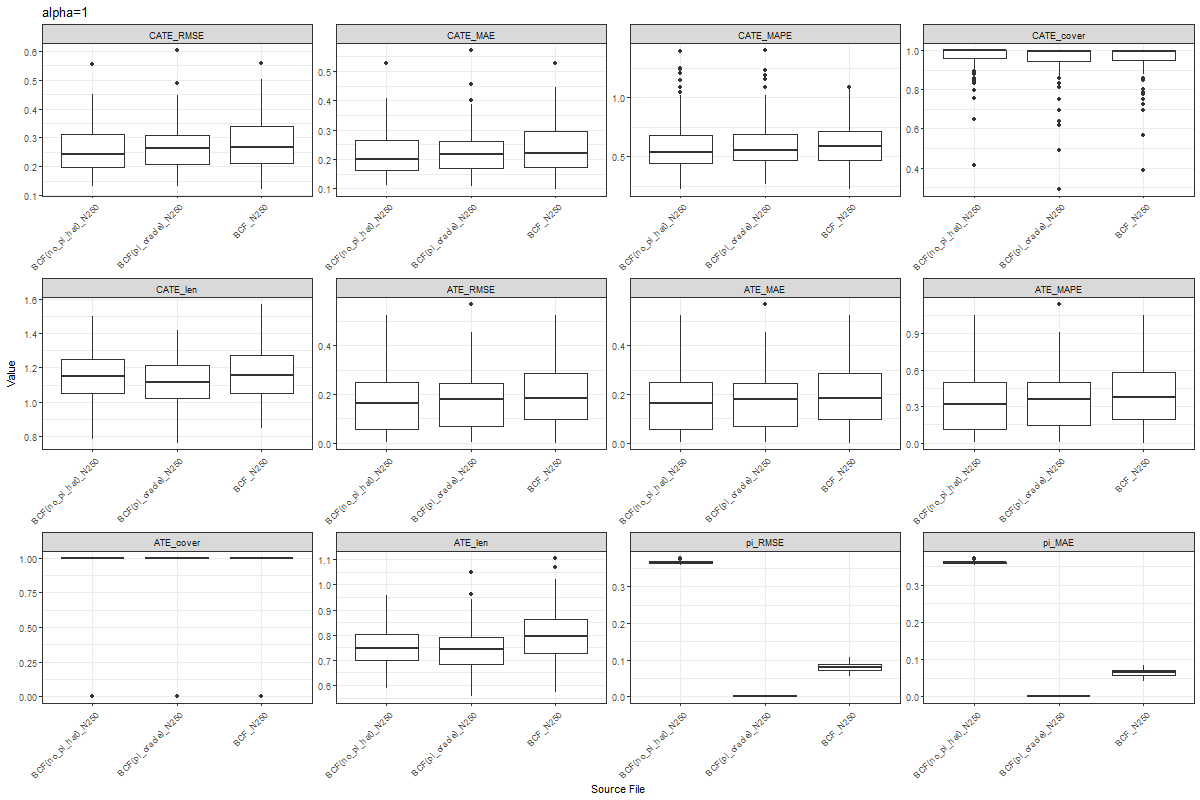}
  \caption{Box Plots for $\alpha=1$}
  \label{Figure5}
  \centering
  \end{figure}
  
  \begin{figure}[H]
  \centering
  \includegraphics[scale=0.5]{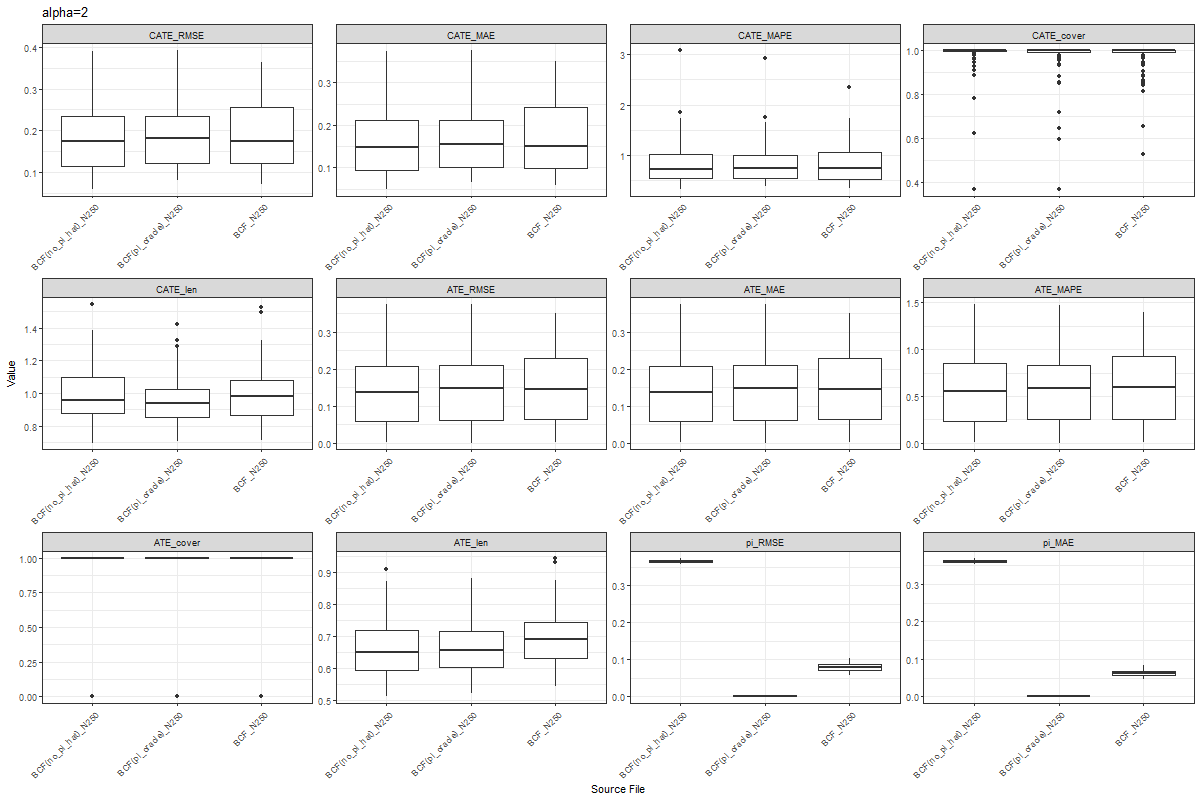}
  \caption{Box Plots for $\alpha=2$}
  \label{Figure6}
  \centering
  \end{figure}
  
  \begin{figure}[H]
  \centering
  \includegraphics[scale=0.5]{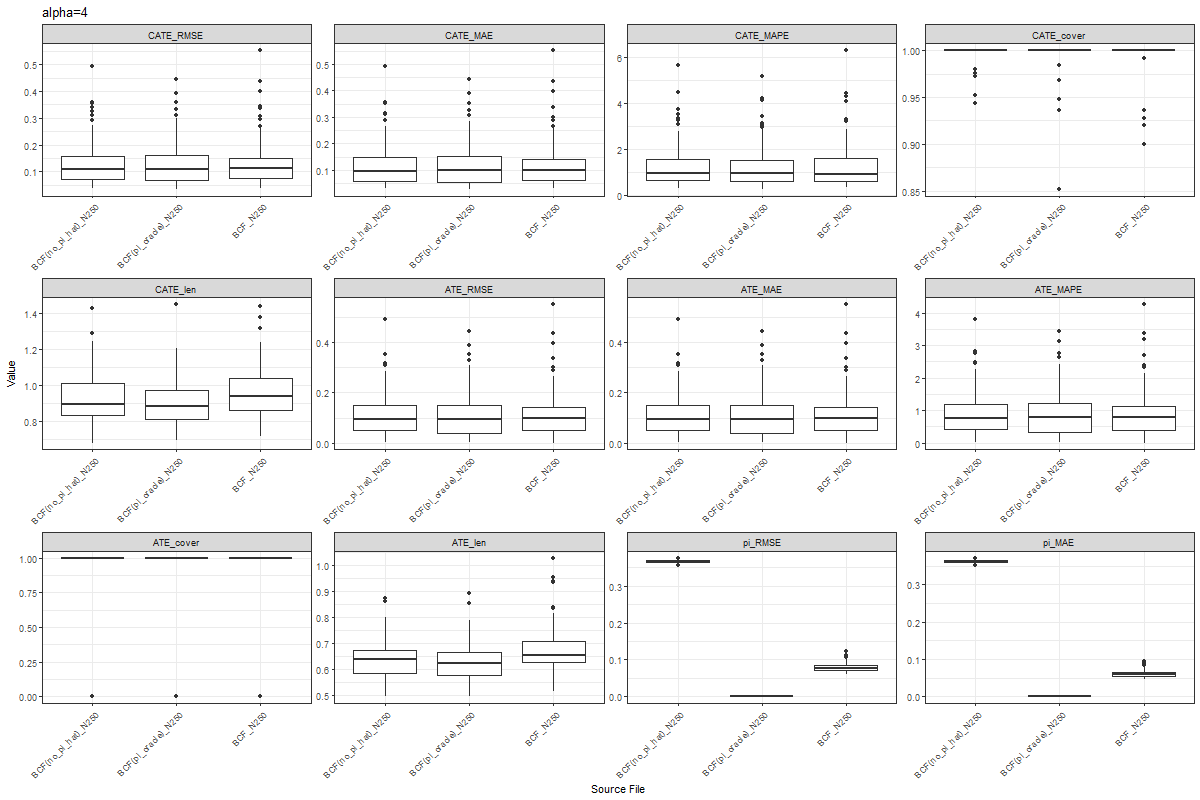}
  \caption{Box Plots for $\alpha=4$}
  \label{Figure7}
  \centering
  \end{figure}

  \subsection{DGP3}
  \begin{figure}[H]
  \centering
  \includegraphics[scale=0.5]{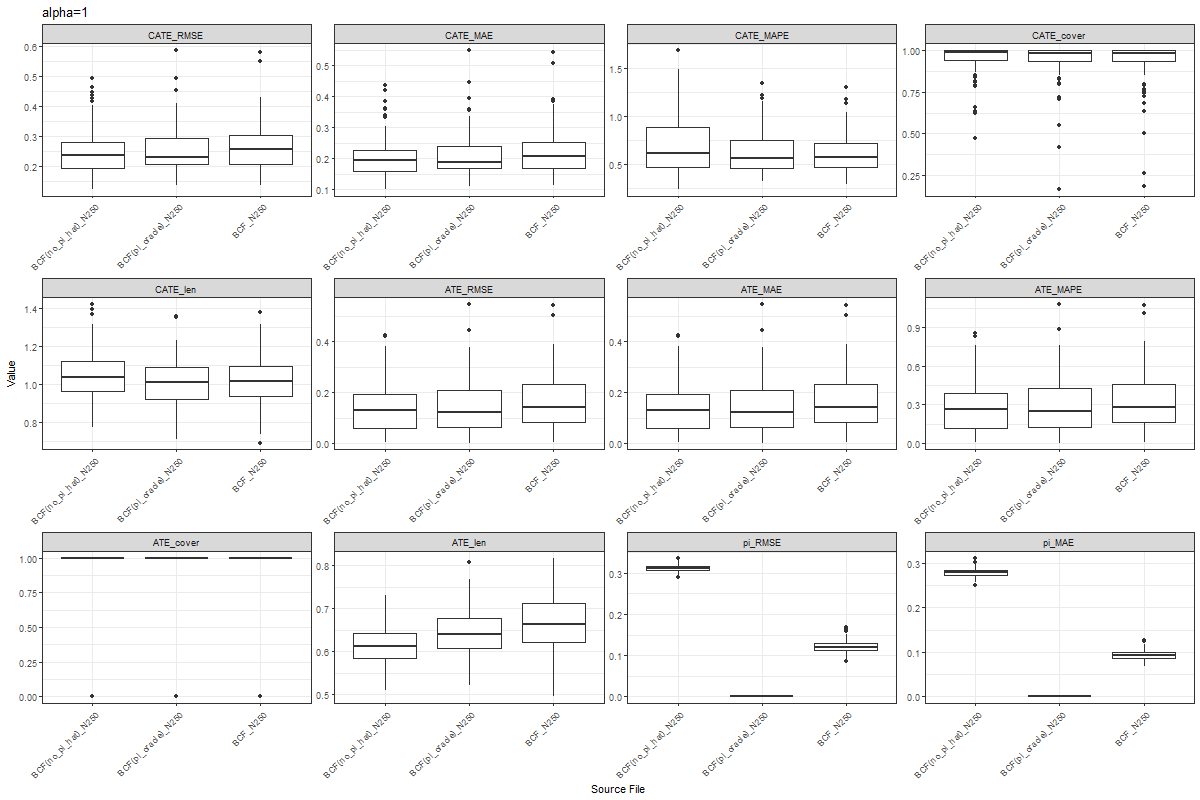}
  \caption{Box Plots for $\alpha=1$}
  \label{Figure8}
  \centering
  \end{figure}
  
  \begin{figure}[H]
  \centering
  \includegraphics[scale=0.5]{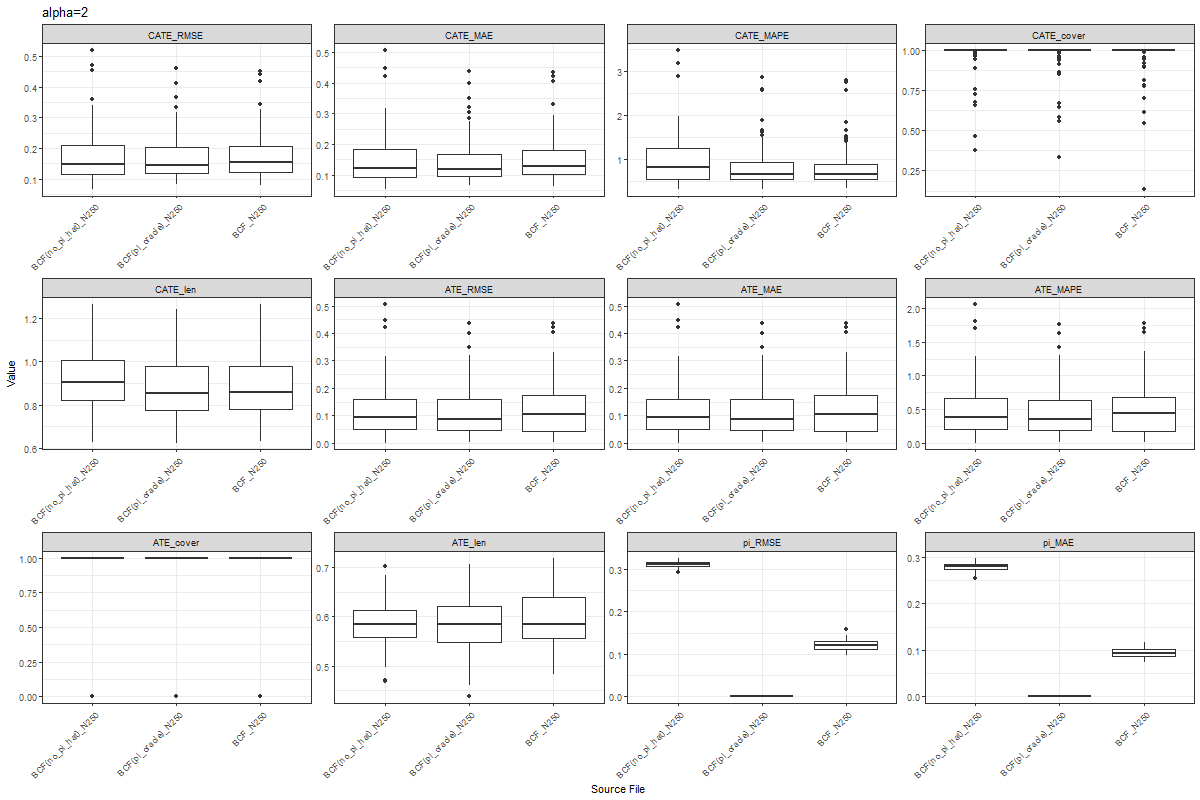}
  \caption{Box Plots for $\alpha=2$}
  \label{Figure9}
  \centering
  \end{figure}
  
  \begin{figure}[H]
  \centering
  \includegraphics[scale=0.5]{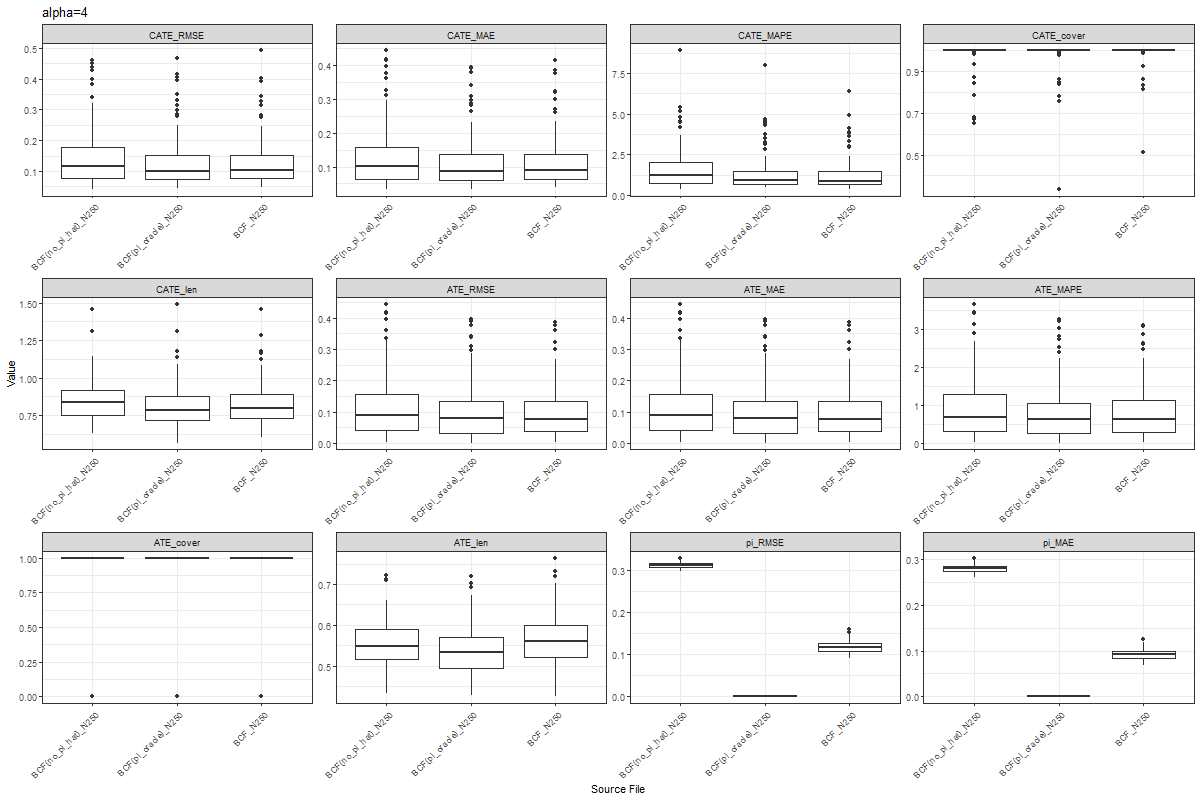}
  \caption{Box Plots for $\alpha=4$}
  \label{Figure10}
  \centering
  \end{figure}

\end{landscape}

\end{document}